\documentclass[acmlarge, prologue, table]{acmart}
\usepackage[title]{appendix}
\usepackage{tabularx}
\usepackage{graphicx}
\usepackage{adjustbox}
\usepackage{listings}
\usepackage[normalem]{ulem}
\usepackage{tcolorbox}
\usepackage{pdflscape}
\usepackage{hyperref}
\usepackage{multicol}
\usepackage{multirow}
\usepackage{subcaption}
\usepackage{tcolorbox}
\usepackage{float}
\usepackage{makecell}

\newbox{\picbox}

\AtBeginDocument{%
  \providecommand\BibTeX{{%
    \normalfont B\kern-0.5em{\scshape i\kern-0.25em b}\kern-0.8em\TeX}}}

\setcopyright{rightsretained}
\acmJournal{IMWUT}
\acmYear{2024} \acmVolume{8} \acmNumber{1} \acmArticle{12} \acmMonth{3} \acmPrice{}\acmDOI{10.1145/3643505}

\newcommand\jsonkey{\color{purple}}
\newcommand\jsonvalue{\color{cyan}}
\newcommand\jsonnumber{\color{orange}}

\newcommand{\aeone}[1]{{\color{black} #1}}
\newcommand{\aetwo}[1]{{\color{black} #1}}
\newcommand{\rthree}[1]{{\color{black} #1}}
\newcommand{\rfour}[1]{{\color{black} #1}}
\newcommand{\changed}[1]{{\color{black} #1}}

\makeatletter
\newif\ifisvalue@json

\lstdefinelanguage{json}{
    tabsize             = 4,
    showstringspaces    = false,
    keywords            = {false,true},
    alsoletter          = 0123456789.,
    morestring          = [s]{"}{"},
    stringstyle         = \jsonkey\ifisvalue@json\jsonvalue\fi,
    MoreSelectCharTable = \lst@DefSaveDef{`:}\colon@json{\enterMode@json},
    MoreSelectCharTable = \lst@DefSaveDef{`,}\comma@json{\exitMode@json{\comma@json}},
    MoreSelectCharTable = \lst@DefSaveDef{`\{}\bracket@json{\exitMode@json{\bracket@json}},
    basicstyle          = \ttfamily
}

\newcommand\enterMode@json{%
    \colon@json%
    \ifnum\lst@mode=\lst@Pmode%
        \global\isvalue@jsontrue%
    \fi
}

\newcommand\exitMode@json[1]{#1\global\isvalue@jsonfalse}

\lst@AddToHook{Output}{%
    \ifisvalue@json%
        \ifnum\lst@mode=\lst@Pmode%
            \def\lst@thestyle{\jsonnumber}%
        \fi
    \fi
    \lsthk@DetectKeywords%
}

\begin{document}

\title{Sasha: Creative Goal-Oriented Reasoning in Smart Homes with Large Language Models}

\author{\href{https://orcid.org/0000-0003-4689-4591}{Evan King}}
\affiliation{%
  \institution{University of Texas at Austin}
  \city{Austin}
  \state{TX}
  \country{USA}}
\email{e.king@utexas.edu}

\author{\href{https://orcid.org/0000-0002-3518-946X}{Haoxiang Yu}}
\affiliation{%
  \institution{University of Texas at Austin}
  \city{Austin}
  \state{TX}
  \country{USA}}
\email{hxyu@utexas.edu}

\author{\href{https://orcid.org/0000-0002-3348-3163}{Sangsu Lee}}
\affiliation{%
  \institution{University of Texas at Austin}
  \city{Austin}
  \state{TX}
  \country{USA}}
\email{sethlee@utexas.edu}

\author{\href{https://orcid.org/0000-0002-4131-4642}{Christine Julien}}
\affiliation{%
  \institution{University of Texas at Austin}
  \city{Austin}
  \state{TX}
  \country{USA}}
\email{c.julien@utexas.edu}
\renewcommand{\shortauthors}{E. King, et al.}
\pagenumbering{arabic}
\begin{abstract}
Smart home assistants function best when user commands are direct and well-specified---e.g., ``turn on the kitchen light''---or when a hard-coded routine specifies the response. In more natural communication, however, human speech is unconstrained, often describing goals (e.g., ``make it cozy in here'' or ``help me save energy'') rather than indicating specific target devices and actions to take on those devices. Current systems fail to understand these under-specified commands since they cannot reason about devices and settings as they relate to human situations. We introduce large language models (LLMs) to this problem space, exploring their use for controlling devices and creating automation routines in response to under-specified user commands in smart homes. We empirically study the baseline quality and failure modes of LLM-created action plans with a survey of age-diverse users. We find that LLMs can reason creatively to achieve challenging goals, but they experience patterns of failure that diminish their usefulness. We address these gaps with Sasha, a smarter smart home assistant. Sasha responds to loosely-constrained commands like ``make it cozy'' or ``help me sleep better'' by executing plans to achieve user goals---e.g., setting a mood with available devices, or devising automation routines. We implement and evaluate Sasha in a hands-on user study, showing the capabilities and limitations of LLM-driven smart homes when faced with unconstrained user-generated scenarios.
\end{abstract}




\maketitle

\section{Introduction}
\label{introduction}
A long-standing challenge in ubiquitous computing is to develop smart spaces that can \emph{creatively infer and respond to the goals of users}. A person who speaks conversationally to a smart home interface like Alexa---e.g., asking it to ``make the living room cozy''---quickly discovers that these systems struggle to relate everyday human situations to appropriate smart home actions~\cite{kim2021exploring, pradhan2020use, cowan2017can, luger2016like}. When people interact with one another, 
unconstrained natural language is less of a problem. If a person asks us to ``make the living room cozy'', we 
relate the concept of ``coziness'' to 
qualities of the environment: 
temperature, 
lighting, 
music, etc. The right response varies based on the space, the parameters the space provides, and individual preferences~\cite{reisinger2022user, jeong2010smart}. 
Generally speaking, humans are 
good at making quick semantic leaps from abstract concepts like ``cozy'' to specific actions to take on the available devices.
The systems that drive modern-day smart homes unfortunately lack this ability.

When users discover the limits of home assistants' reasoning, they 
must manually install smart home routines. These tools (e.g., ``If This Then That'' (IFTTT)~\cite{ifttt2023}) provide a way to imbue the modern smart home with the \emph{illusion} of intelligence, though the true source of intelligence in this case is still the \emph{user} who conjures it. 
If someone wants to tell Alexa ``it's party time'' in order to trigger a lighting scheme and stereo configuration, they must use IFTTT to hard-code the action. 
By a recent count, around 3,000 smart home routines from IFTTT have been installed a total of nearly 2 million times~\cite{yu2021analysis}, which implies that \emph{users have many goals that are not achieved by current systems without manual intervention.} Users must 
adapt their mental model to the system's limitations, rather than the system intuitively supporting the user~\cite{upadhyay2023studying, clark2017devices}.

We approach the challenge of flexibly supporting users' goals in smart homes using large language models (LLMs). LLMs are general-purpose language models trained on diverse corpora that span much of the written text and code available on the internet~\cite{radford2019language, gao2020pile}. These models have shown remarkably high performance on many downstream tasks without a need for major model updates tailored to specific use-cases~\cite{brown2020language, wei2021finetuned}. The same GPT-3 model, for instance, has been used to control robotics with natural language~\cite{liang2022code, wu2023tidybot} and to simulate human social dynamics in small communities~\cite{park2023generative}. 
Some of this generalizability 
is attributed to the diversity and scale of LLMs' training data: since this data spans disciplines and contexts, it embodies cross-cutting semantic relationships in a way that data input to task-specific models does not~\cite{gao2020pile, touvron2023llama}. As Dey wrote in a foundational paper on context-aware systems, it is ``the richness of [human] language... [a] common understanding of how the world works, and an implicit understanding of everyday situations'' that enables humans to communicate in terms of high-level ideas rather than specific actions~\cite{dey2001understanding}. Given LLMs' performance at many tasks that involve unconstrained language and human situational contexts, we introduce their use in smart environments. We investigate their potential to \emph{bridge the conceptual gap between implicit smart home user goals and specific actions needed to accomplish them,} \aeone{structuring our inquiry around these research questions:

\begin{itemize}
    \item \emph{\textbf{RQ1:} \label{rq1} What unique capabilities are unlocked when LLMs are used for smart home control?} We seek to uncover if and how LLMs can provide support for more loosely-constrained interactions in smart homes.
    \item \emph{\textbf{RQ2:} \label{rq2} What practical challenges will LLM-based systems present?} LLMs can produce incorrect or ``hallucinated'' responses and are computationally expensive to run. We seek to establish a baseline understanding of these practical issues as they exist in smart homes.
    \item \emph{\textbf{RQ3:} \label{rq3} What system design choices can address these practical challenges?} We seek to apply LLM system designs from other domains and gauge their usefulness and suitability for tackling the application-specific challenges of smart homes.
    \item \emph{\textbf{RQ4:} \label{rq4} How well can this new form of smart home support user goals in unconstrained scenarios?} In a real-world implementation, we seek to understand how users interact with a less-constrained system, what new capabilities it provides, and what limitations still exist.
\end{itemize}

\begin{figure}
    \centering
    \includegraphics[width=\textwidth]{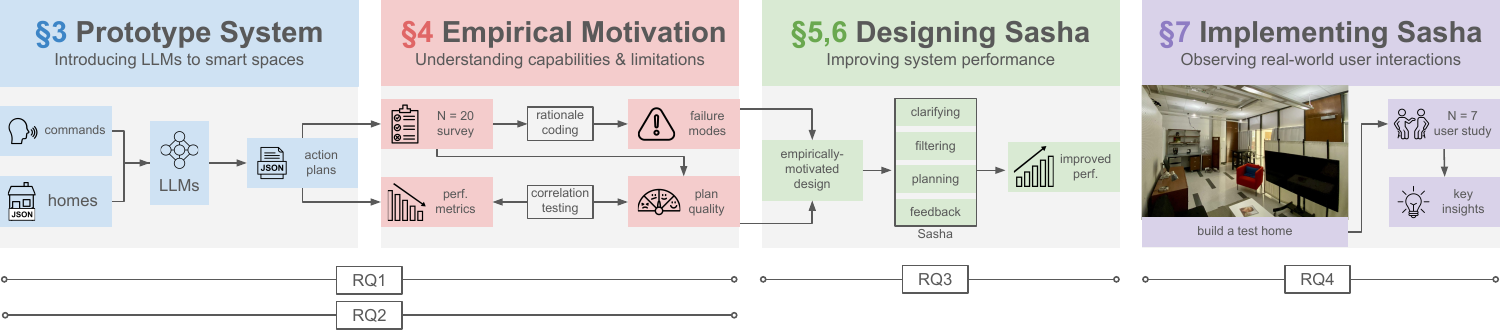}
    \caption{Overview of the paper's four key components, their internal structure, and motivating research questions.}
    \label{fig:paper-overview}
\end{figure}

\noindent
We investigate \textbf{RQ1-4} throughout this paper, which consists of the following four efforts (summarized in Fig.~\ref{fig:paper-overview}).}

\changed{\textbf{Introducing LLMs to smart spaces.} There is little prior work at the intersection of LLMs and smart environments. We introduce LLMs to the problem space with a prototype system that responds to user commands in smart homes with concrete, executable \emph{action plans} that make use of available devices and sensors.}
 
\textbf{Empirical motivation.} We study the outputs of our prototype system to establish an empirical understanding of the 
challenges that arise when LLMs are used in smart environments. We qualitatively analyze 600 labels and free-form rationales from $N=20$ human annotators regarding LLM-generated action plans, supplemented by a quantitative analysis. We find that LLMs can flexibly generate creative action plans in response to under-specified user commands. LLMs are also prone to failures at the task, however; we reveal 7 common types of failure responsible for low user satisfaction and propose methods for measuring them. We show that LLMs can realize \emph{immediate user goals} like ``make it cozy in here'', as well as more complex \emph{persistent goals} like ``use natural light when the sun is out'' that require automation routines.

\textbf{Designing Sasha.} Toward addressing the challenges that we identify, we propose Sasha: a \underline{s}m\underline{a}rter \underline{s}mart \underline{h}ome \underline{a}ssistant. 
Sasha's novelty lies in an  \emph{iterative reasoning process} that enables it to work creatively to achieve user goals while modulating LLMs' tendency to produce non-sensical or hallucinated plans. Our design addresses the challenges of \emph{targeting relevant devices} and \emph{incorporating user preferences} that we empirically motivate. Our evaluations show that Sasha reduces false positives and erroneous targeting of unrelated devices.

\textbf{Implementing Sasha.} We implement Sasha in a test home. \changed{We invite $N=7$ participants to give unconstrained commands 
during user-generated scenarios, 
allowing us to gather user perspectives and measurements of the system's performance in a realistic setting.}

In summary, our contributions are as follows:

\begin{itemize}
    \item \changed{A system that introduces LLMs to the problem of reasoning in response to user commands in smart spaces.}

    \item An empirical study that demonstrates the potential for LLMs to support goal-oriented interactions, while also revealing specific patterns of failure that diminish LLMs' usefulness.
    
    \item Sasha, a \underline{s}m\underline{a}rter \underline{s}mart \underline{h}ome \underline{a}ssistant that leverages a novel iterative reasoning process to produce high-quality action plans in response to under-specified user commands in different smart homes.
    
    \item A set of evaluations of Sasha that qualitatively and quantitatively validate each step of the iterative reasoning process that the system uses to produce action plans.
    
    \item A real-world implementation and user study of Sasha in a test home that demonstrates the system's practical capabilities when faced with unconstrained, user-generated scenarios.
\end{itemize}

\changed{We provide background in Section~\ref{related-work}. We introduce LLMs to smart environments with a prototype system in Section~\ref{feasibility-study}. We conduct an empirical study in Section~\ref{sec:empirical-motivation} that reveals specific avenues for further research. Section~\ref{approach} introduces Sasha, a system that addresses some key identified challenges. We evaluate Sasha in Section~\ref{evaluation} and describe the results of a real-world user study in Section~\ref{user-study}. Section~\ref{discussion} discusses implications and future work, while Section~\ref{conclusion} concludes.}

\section{Background \& Related Work}
\label{related-work}
In the following, we situate the paper with related research in smart environments and large language models.

\subsection{Smart environments \& user goals}
``Smart environments'' align with Weiser's vision for ubiquitous computing, where computing capabilities embedded in everyday devices are used to unobtrusively meet peoples' needs~\cite{weiser1999computer}. A smart environment relies on networked devices and sensors embedded in light bulbs, 
appliances, wearables, and the built environment to sense and respond to these needs~\cite{dunne2021survey}. Using some form of ``intelligence'' in the built environment to coordinate devices and sensors has broad applications in building management~\cite{kim2022design}, infrastructure~\cite{branny2022smarter}, healthcare~\cite{acampora2013survey}, and homes~\cite{alaa2017review}. We focus 
on smart homes, which Lutolf described as ``the integration of different services within a home'' that ``includes a high degree of \ldots flexibility''~\cite{lutolf1992}. We consider smart homes from a socio-technical perspective (i.e., how do they support \emph{people}), rather than from a strictly technical perspective (i.e., what can they do), since this ultimately influences whether 
people choose to incorporate them in their daily lives~\cite{wilson2015smart}. In a recent socio-technical analysis of smart homes, 
Reisinger et al. identify customizability, automation, accessibility, reliability, and low latency as primary concerns~\cite{reisinger2022user}. Each of these influences our approach.

Our work centers on the challenge of \emph{reasoning} about available devices and sensors in a home in relation to the implicit \emph{goals} expressed by a user in their own manner of speaking~\cite{clark2017devices, noura2020vish, palanca2018designing}. We frame the problem in terms of creating executable ``action plans'' that use the space's capabilities to meet a user's \emph{immediate} (i.e., doing something right now) and \emph{persistent} (i.e., automating something) goals. Our separation of interactions into these two categories mirrors that of Clark et al., who 
classify interaction patterns as either immediate or ``conditional''~\cite{clark2017devices}. That study found that 
users ``expect agents to understand basic \emph{goals.} If the system were able to support this, it would provide the agent with a great deal of flexibility in achieving the specified goals.'' In these envisioned goal-oriented interactions, systems should be made to infer and devise courses of action in response to users' descriptions of desired end states or changes to the environment (e.g., ``I want to watch a movie'') rather than descriptions of specific actions~\cite{georgievski2016automated}. 

\begin{figure}
    \centering
    \includegraphics[width=\textwidth]{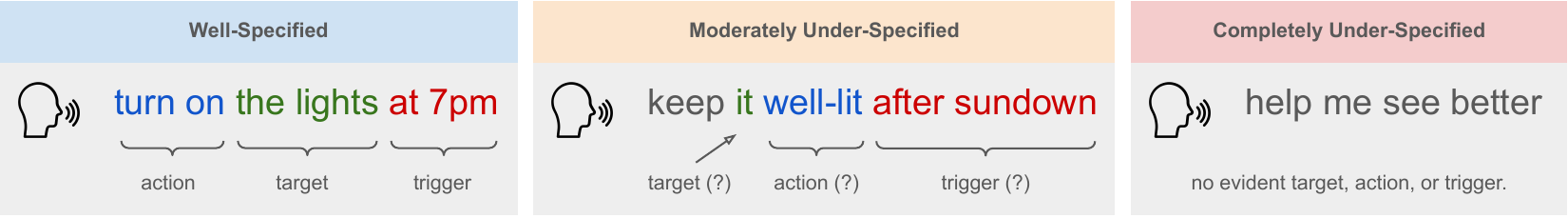}
    \caption{\aeone{User commands have varying degrees of specificity. Left: Well-specified commands define specific actions, specific target devices, and specific triggers. Middle: Moderately under-specified commands may allude to targets, actions, and triggers, but do not specify them. Right: Completely under-specified commands have no obvious targets, desired actions, or triggers. We focus on moderately and completely under-specified commands.}}
    \label{fig:under-specified}
\end{figure}

\label{rev:1ae-1}\aeone{We focus on user commands with goals that are \emph{under-specified} in some aspect, i.e., they do not define specific \emph{target devices}, specific \emph{actions} to take on devices, and, in the case of persistent goals, specific \emph{triggers}. The ``specificity'' of a command falls on a spectrum, from well-specified (e.g., ``turn on the kitchen light at 8am'') to moderately under-specified (e.g., ``lighten up the kitchen in the morning'') to completely under-specified (e.g., ``help me wake up'') (Fig.~\ref{fig:under-specified}). In the case of moderately under-specified commands, the goal is often clear to a human but less obvious to a system, which is a common source of frustration: 
existing systems struggle outside of a rigid, highly-specific command structure~\cite{kim2021exploring, cowan2017can, luger2016like} and are generally unable to deal with complex commands~\cite{upadhyay2023studying, pradhan2020use}. For this reason, our scope includes moderately under-specified commands (e.g., ``turn the AC off when it's cold outside'') along with completely under-specified commands (e.g., ``make it cozy in here'').}

Prior work has proposed 
task-specific 
systems 
for inferring user goals. Noura et al. proposed the VISH grammar and generative dataset for modeling goal-oriented smart home utterances, using it to train task-specific natural language models and improve understanding~\cite{noura2020natural, noura2020vish}. This approach broadens the supported command structure of assistants by introducing a goal-oriented grammar, e.g., a user can say ``the blinds are too low'' to raise the blinds. The system does not provide support for completely under-specified commands like ``make it cozy in here'' that might entail 
different responses depending on the environment, nor does it provide support for persistent goals like ``help me save energy'' that require reasoning about a home's devices and sensors. Palanca et al. 
approach the problem of achieving user goals with a task-specific system, proposing a multi-agent approach~\cite{palanca2018designing}. Agents (e.g., a lighting agent, a TV agent, etc.) coordinate to form a suitable plan to achieve a user goal based on an ontology that describes the agents' capabilities and relationships. Our focus on output ``action plans'' bears similarity---however, dependence on a rigid ontology ultimately imposes 
assumptions about \emph{how} people use devices in their homes, and what goals they want to accomplish. 
We aim for greater flexibility than existing systems.

\subsection{Applications of large language models}

Large language models (LLMs) come from 
research in natural language processing (NLP) that began in the 1950s and `60s~\cite{Chomsky+2002}. Research in NLP 
aims to enable computers to understand, interpret, and generate human language in a useful way. A recent 
increase in NLP research activity 
is attributed to the introduction of the transformer model by Vaswani et al.~\cite{vaswani2017attention}. Transformer-based models like Bidirectional Encoder Representations from Transformers (BERT)~\cite{devlin2018bert}, Generative Pre-training Transformer (GPT)~\cite{radford2018improving, radford2019language, brown2020language}, Text-to-text Transfer Transformer (T5)~\cite{raffel2020exploring}, and Large Language Model Meta AI (Llama)~\cite{touvron2023llama, touvron2023llama2} are trained on massive and diverse datasets~\cite{gao2020pile}, enabling them to achieve high performance on 
many tasks involving natural language without a need for task-specific training~\cite{brown2020language, wei2021finetuned}. We focus on several popular transformer models in our empirical study: the commercial GPT-3.5 and GPT-4 models, as well as the Llama 2 model. These models have recently been applied to diverse problems in robotics~\cite{wu2023tidybot, skreta2023errors} and social science~\cite{park2023generative}.

In 
smart homes, LLMs 
may unlock more 
flexible interactions between users and spaces than task-specific systems. This stems from a \emph{deeper exposure to everyday situations} that are embodied in their diverse training data. Fast et al., showed, for instance, that a language model trained only on written works of fiction can be used to recognize activities of daily living by leveraging semantic relationships between objects and the activities they are frequently used in~\cite{fast2016augur}. This embodiment of human behavior by language models is effectively a by-product of their training on massive amounts of text that contains unstructured depictions of it. However, with the exception of our own prior small-scale study~\cite{king2023get}, we are not aware of work at the intersection of LLMs and smart spaces. That study showed that GPT-3.5 could devise creative plans in response to under-specified user commands in a smart home. The study did not empirically establish open challenges, 
nor did it investigate the possibility of achieving persistent user goals, propose 
system improvements, or include a user study. We consider all four.

Integrating LLMs into 
systems remains highly subjective~\cite{qiao2022reasoning, shi2023large}. Two overlapping practices exist to adapt general-purpose models to specific tasks: \emph{fine tuning}~\cite{ziegler2019fine} and \emph{prompt engineering}~\cite{liu2023pre}. The former involves partially retraining the model using (input, output) pairs that reflect the desired structure of output, while the latter relies on carefully-constructed natural language prompts that elicit the desired output. In some cases, engineered prompts may contain ``few-shot'' training examples that hint at the desired output structure, while prompts that contain \emph{no} training examples are referred to as ``zero-shot''. Fine tuning and prompt engineering are not mutually exclusive: generally, prompt engineering performs better on models that have already been fine-tuned for \emph{instruction following}~\cite{wei2021finetuned}, which allows them to 
adhere to instructions 
in the prompt. In our work, we utilize prompt engineering. Work with \emph{chain of thought prompting} 
shows that deconstructing a problem into several steps of reasoning (rather than a single prompt) can improve performance~\rfour{\cite{wei2022chain, wu2022ai}.} \rfour{We design a prototype system based on zero-shot prompting for our initial empirical study, allowing us to characterize LLMs' performance at this new task in the base case.} We then propose an 
improved system, Sasha, that adapts chain-of-thought prompting to smart space applications.

\section{Introducing Large Language Models to Smart Spaces}
\label{feasibility-study}
\begin{figure}[t!]
    \centering
    \includegraphics[width=0.7\columnwidth]{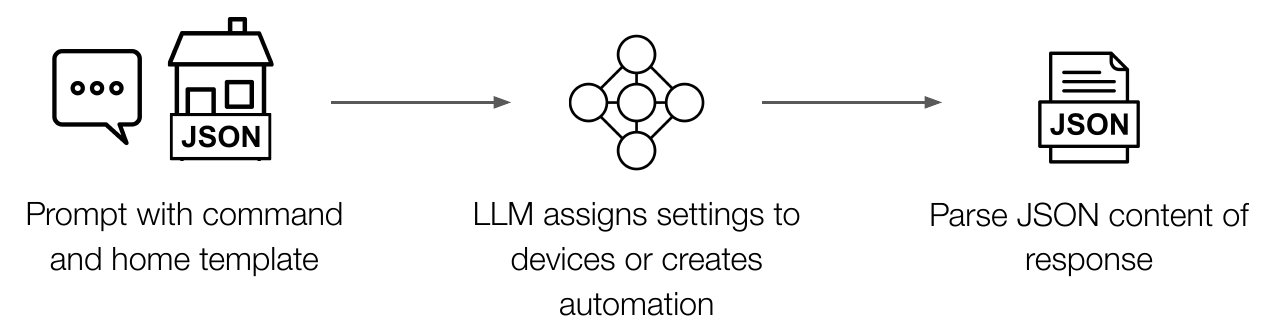}
    \caption{We begin with an initial study of LLM behavior in smart home applications using this experimental setup.}
    \label{fig:prototype}
\end{figure}

We build a prototype 
that utilizes 
LLMs to produce action plans in response to user commands in smart homes. Our prototype works using zero-shot prompts to an LLM containing a user's natural language command alongside a ``home template'' in JSON that describes the rooms, devices, and sensors in a smart home. Instructions in the prompt guide the LLM to output a machine-parseable action plan---i.e., changes to device settings for immediate commands and (trigger, action) pairs for automation routines, both also expressed in JSON---\label{rev:4r-3.1}\rfour{or to reject the user's request if the goal appears unattainable with the devices available}. The JSON action plans generated by the model are 
then 
parsed and executed. \rfour{Since LLMs have 
not been rigorously studied in the context of smart environments, our prototype design avoids venturing 
into state-of-the-art methods 
from other 
domains (e.g., fine-tuning or chain-of-thought prompting). This provides an understanding of \emph{performance in the base case} along with a baseline for future improvements.} We propose such improvements later in Section~\ref{approach}. We illustrate our setup in Fig.~\ref{fig:prototype} and provide detail in the following.

\begin{figure}
    \centering
    \includegraphics[width=\columnwidth]{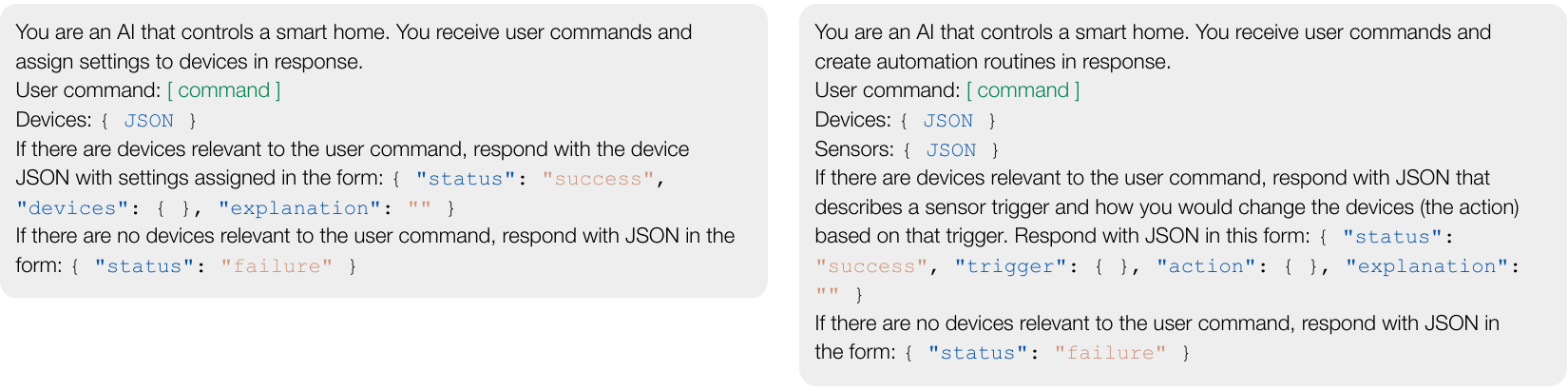}
    \caption{The text of immediate and persistent goal prompts (left and right, respectively). Immediate prompts produce action plans that can be executed immediately, while persistent prompts produce automation routines.}
    \label{fig:prompts}
\end{figure}

\textbf{Prompting.} Our prompts contain (1)~instructions about the task, (2)~a JSON ``home template'' containing information about the home's rooms, devices, sensors, and settings, and (3)~the user's command. Prompts for immediate commands instruct the model to output an action plan to execute immediately, while persistent command prompts instruct the model to output an automation routine in the form of a (trigger, action) pair. The trigger is a sensor or set of sensors and the corresponding values that 
trigger the routine, while the action is a set of devices and the settings 
to assign to them. In both prompt types, we instruct the model to reject the user's command if the goal appears unachievable with the devices available. We illustrate the prompts in Fig.~\ref{fig:prompts}.

\begin{figure}
    \centering
    \includegraphics[width=\columnwidth]{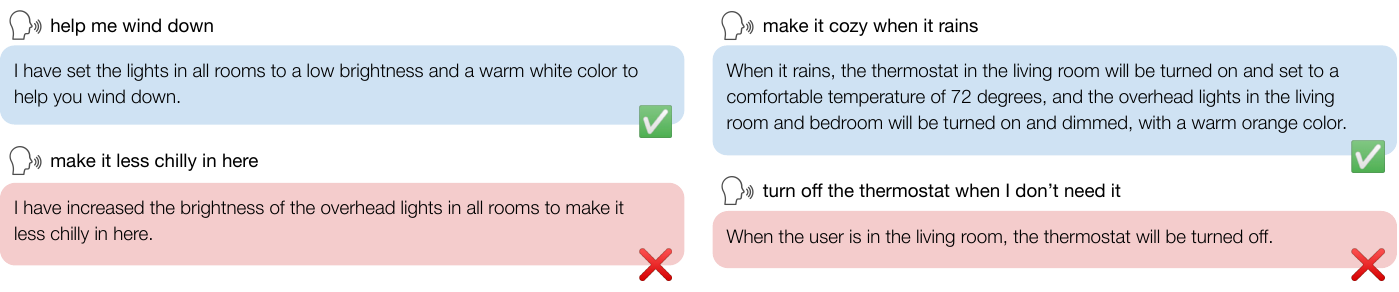}
    \caption{GPT-3.5 responses to smart home commands. We prompt the model with a user command along with information about the devices and sensors available in the home. Responses are often remarkably creative, and often remarkably wrong.}
    \label{fig:example-responses}
\end{figure}

\textbf{Planning.} Given the prompt, the LLM constructs a JSON action plan that suggests a mapping of the goal inferred from the user's command to specific device settings that accomplish the goal. Since this ``reasoning'' ultimately stems from the semantic relationships between different words, the action plan is constructed based on the \emph{naming} of rooms, devices, and settings in the JSON of the home template. A device in the template named \texttt{lamp} expresses both the \emph{functional capabilities} of a lamp (e.g., luminance) as well as the \emph{semantic content} of \texttt{lamp} (e.g., lamps are used when reading, lamps provide mood lighting). 
This motivates the descriptive naming and structuring of the home template, since it is the primary means by which the specific (e.g., setting) and abstract (e.g., ``mood'') capabilities of the home are made available to the model.

\textbf{Parsing.} After the LLM has produced an action plan, we post-process it to extract the JSON. Our prompts contain instructions to include an ``explanation'' of the changes made in the JSON for easier inspection. With this system in place, an LLM can respond to smart home commands---some examples are illustrated in Fig.~\ref{fig:example-responses}.

\section{Empirical Motivation}
\label{sec:empirical-motivation}Toward answering \textbf{RQ1-2}, we conduct an empirical study using the prototype system described previously. We (1)~construct a representative dataset of user commands and smart home configurations to use as input to our system, (2)~design a survey to elicit user perspectives on the quality and modes of failure of the system's action plans, (3)~establish quantitative metrics on the action plans themselves, (4)~and report our findings for both the survey and the quantitative analysis.

\subsection{Home templates \& user commands}
\label{sec:templates-and-commands}
\changed{We construct a representative dataset to use as input to our system, allowing us to produce a large set of action plans for analysis. We describe our method for constructing ``home templates'' to use as input to the system, then describe our procedure for creating user commands that are representative of our task.} 

\textbf{Home templates}. 
We 
constructed three home templates by adapting data from two sources: a dataset of IFTTT smart home routines~\cite{yu2021analysis} and the layout and sensor suite from a home in the CASAs dataset~\cite{cook2012learning}. The IFTTT data consists of the most popular ``recipes''  users have created and installed to control devices and automate routines. It indicates \emph{which} devices people are using and \emph{how} they are using them, as well as types of goals that are not met by current systems without manual configuration. 

\begin{figure}[t!]
    \centering
    \includegraphics[width=0.3\columnwidth]{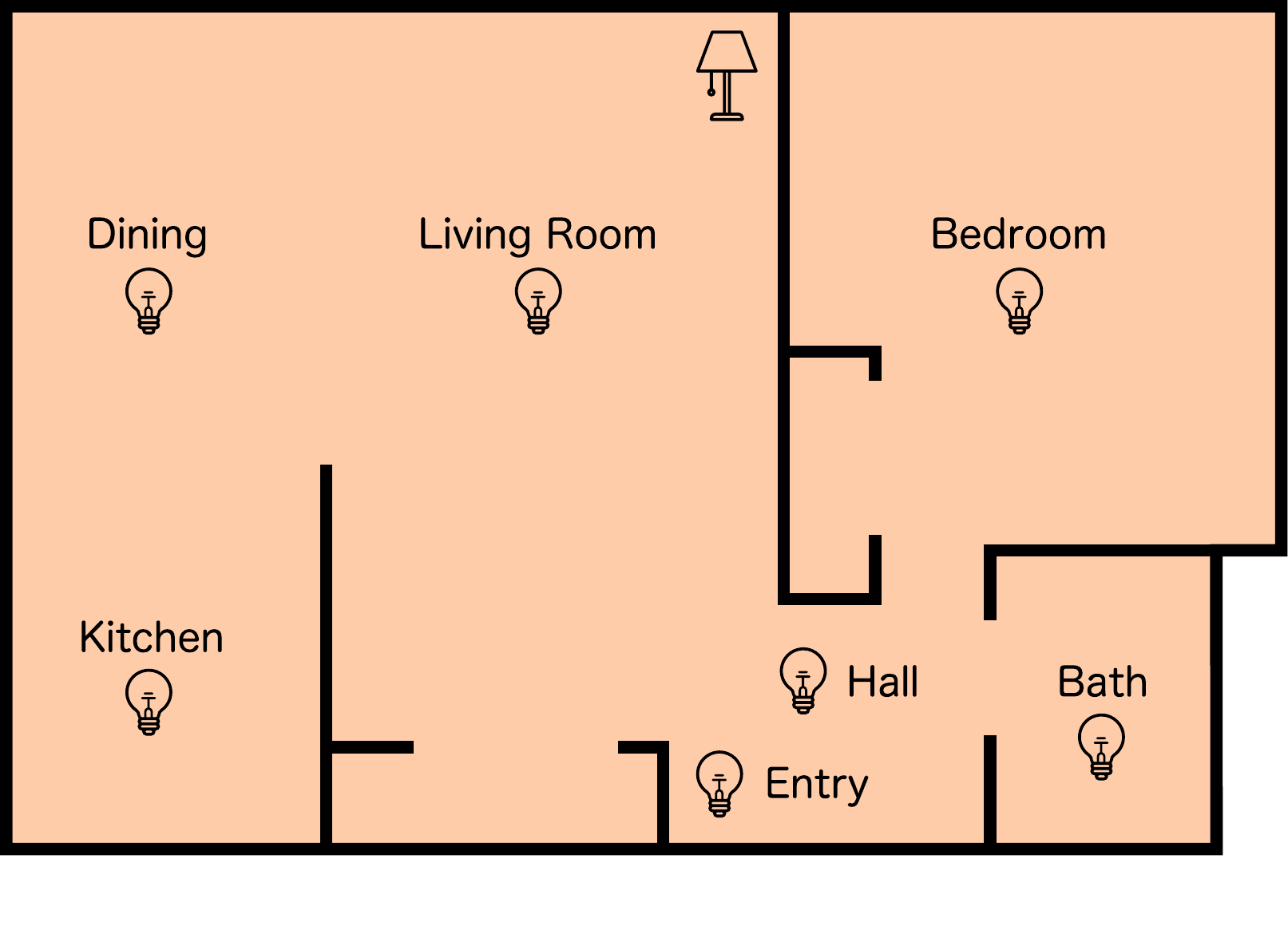} \hfill
    \includegraphics[width=0.3\columnwidth]{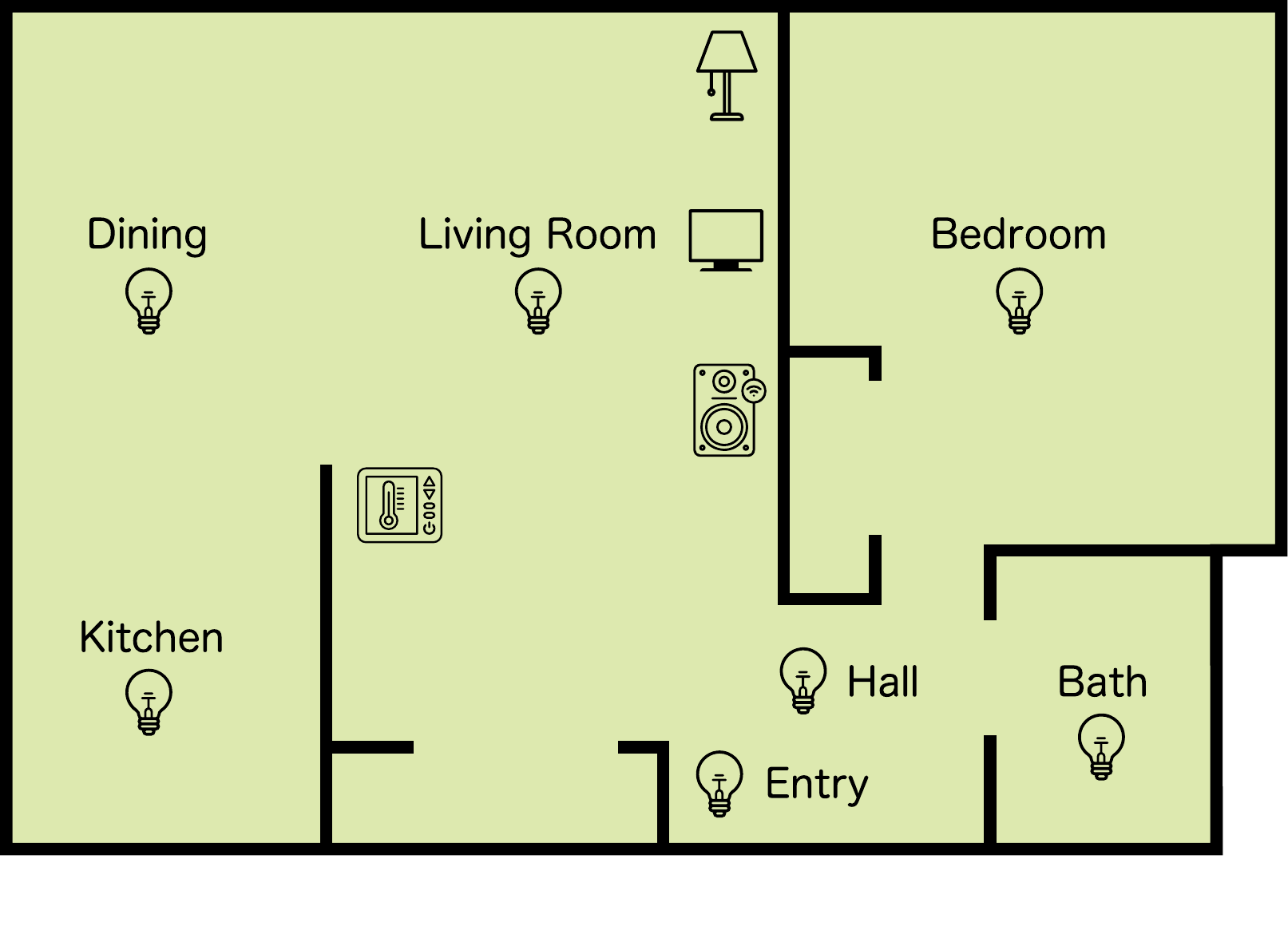} \hfill
    \includegraphics[width=0.3\columnwidth]{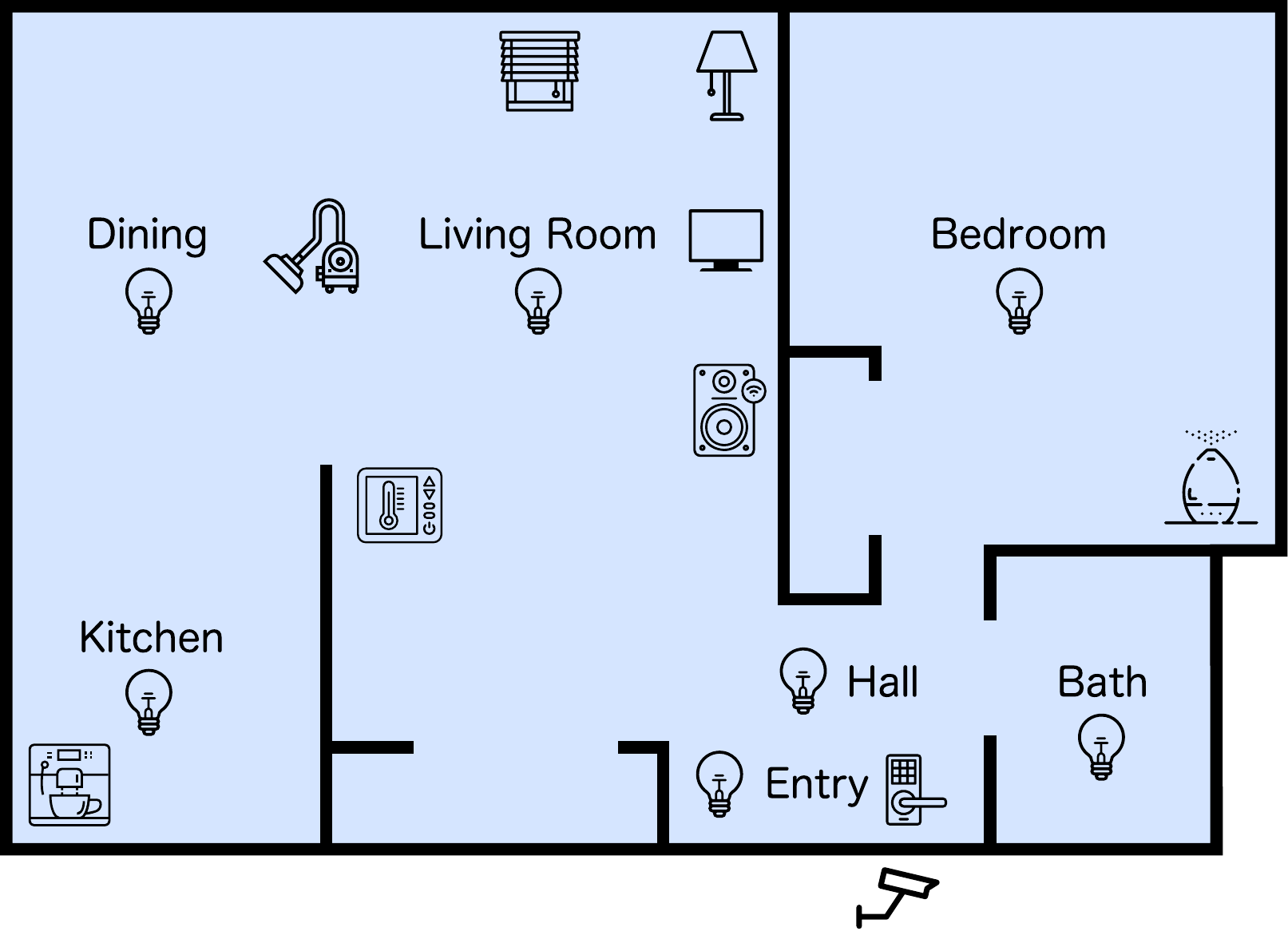}
    \caption{We model three homes ($h_1$, $h_2$, and $h_3$ from left to right) with an increasingly diverse set of \emph{devices}. All three homes have the same suite of \emph{sensors}. The model's reasoning is based on a given command and the devices and sensors available.}
    \label{fig:home-templates}
\end{figure}

\begin{figure}[t!]
    \centering
    \includegraphics[width=0.5\columnwidth]{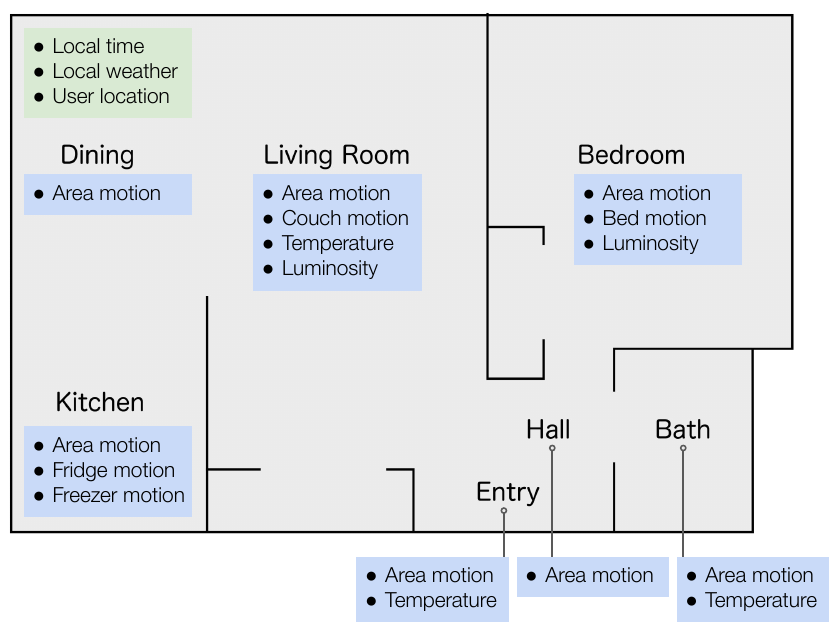}
    \caption{Each home template in our set of homes $h_1$, $h_2$, and $h_3$ has the same suite of sensors, as shown.}
    \label{fig:sensors}
\end{figure}

We modeled three homes, $h_1, h_2$, and $h_3$, with different number and diversity of devices. We chose which device types to include in each home based on their 
popularity in the IFTTT dataset. Smart lights are 
the most popular, 
so $h_1$ models the most basic smart home with only lights. $h_2$ adds the next most popular types (climate control and entertainment), while $h_3$ again adds the next most popular types (security, a robot vacuum, and miscellaneous appliances). Each home has the same suite of sensors, built by taking a subset of the sensors in home HH101 in the CASAs dataset. Each home has the same layout as HH101---we illustrate the homes and devices in Fig.~\ref{fig:home-templates} and the sensor suite in Fig.~\ref{fig:sensors}.

We encode each home in a JSON ``home template'' using two data structures: one for the home's controllable \texttt{devices} and another for its \texttt{sensors}. Each top-level element of the home is a room, named \emph{descriptively} (e.g., \texttt{livingroom}) as a semantic cue about 
the activities that take place there. Rooms contain devices, again named descriptively (e.g., \texttt{overhead\_light}). 
Each device has several settings, expressed as key-value pairs, where the key is the name of the setting (e.g., a light's \texttt{brightness}) and the value is its data type (e.g., \texttt{float}). A second structure 
contains the \texttt{sensors} in the home---e.g., motion, luminosity---also organized by room. We include a \texttt{global} field in the sensor structure for values like \texttt{local\_time} and \texttt{weather}. We choose JSON since it is 
common in LLM training corpora 
~\cite{touvron2023llama, gao2020pile}. It is furthermore the de-facto format for smart device APIs~\cite{hue2022, kasa2023, nest2022, insteon2022}, which eases the transition from research to implementation.

\textbf{User commands.} We created 40 natural language commands that correspond to common user goals for smart homes. Our commands (1)~reflect user goals not met by current systems and (2)~have challenging, under-specified phrasing. We again mined the IFTTT dataset, examining \emph{how} people use their 
devices. Routines in the dataset are grouped into 12 \emph{goal categories} that describe the types of actions and devices they cover. The ``Ambient Luminance'' category, for instance, includes routines that express a user's goal to change the lighting. We created commands for each category, with the intuition that the existence of a corresponding IFTTT routine implies that manual intervention was required to meet the goal in existing systems. We omit several goal categories that are not relevant to the devices in our homes---Gardening, Device Hubs, Outlets, and Alerts---resulting in 7 categories for 
our commands: Ambient Temperature, Ambient Luminance, Security, Energy Saving, Ambient Atmosphere, Robot Control, and Other Appliances. We created 6 commands per category, with the exception of the Energy Saving category. For this category, we created only 4 commands since many of the energy saving routines in the dataset accomplished similar goals, but simply targeted different manufacturers of similar device types. We 
authored \emph{declarative} commands, i.e., commands that express a desired end state without specifying the required actions~\cite{georgievski2016automated}. \label{rev:1ae-1.2}\aeone{Furthermore, each of our commands is either moderately or completely \emph{under-specified} with respect to at least one of the following: intended \emph{target device(s)}, desired \emph{action}, and, in the case of persistent goals, suitable \emph{trigger} (Fig.~\ref{fig:under-specified}).} \label{rev:3r-1.1}\rthree{The command ``keep it well-lit after sundown'' is an example of a moderately under-specified command since it specifies neither a target (lights, blinds, etc.), nor trigger (a specific time and/or sensor with which to detect it).} ``Help me lower my power bill'' is an example of a completely under-specified command since it alludes to no target, trigger, or action that will achieve the goal.

Half of the commands in each category reflect \emph{immediate} goals, and half are \emph{persistent}. Immediate goals 
are achieved through an immediate action or set of actions, e.g., ``make it cozy in here'' or ``make it less stuffy''. Persistent goals 
require automation routines. Examples include ``help me save energy'' or ``let me know if the weather is bad''. We list all commands in Appendix~\ref{appendix-commands} with examples of the IFTTT routines that they are based on.


\subsection{User Survey Procedures \& Metrics}
\label{sec:user-survey-procedures}
We use an online survey to solicit user perspectives on the quality of action plans. We generate action plans using each command and home template combination as input to our experimental setup described previously. We describe survey design, participant recruitment, and analysis procedures in the following.

\subsubsection{Survey design}
\label{sec:user-survey-design}
For a given action plan, we display a picture of the home (as depicted in Fig.~\ref{fig:home-templates}) and the user command given to the home (e.g., ``make it cozy in here''). \label{rev:4r-2.1} We ask participants to label whether each of the \emph{target}, \emph{action}, and (in the case of persistent goal commands) \emph{trigger} of the user command were \emph{well-specified}. We include a definition of well-specified in the survey instructions (Appendix~\ref{appendix-survey}), similar to Fig.~\ref{fig:under-specified}. These labels allow us to check our assumptions about how under-specified users perceive the commands to be. Next, we provide a textual description of the LLM's action plan as extracted from the JSON. We ask participants to label ``how satisfied'' they would be with the actions taken by the smart home on a 5-point scale from ``Very unsatisfied'' to ``Very satisified''. Toward answering \textbf{RQ1-2}, the satisfaction label allows us to analyze possible factors that result in high or low user-perceived quality of the action plans that the LLM generates. Finally, we ask the participant to briefly describe in their own words which aspects of the response were satisfactory or unsatisfactory. The free-form rationale given by the labeller provides important information about both the capabilities and potential failure modes of the LLM. Each survey contains $30$ such questions and takes approximately $45$ minutes to complete. \label{rev:2ae-3.1} Each action plan receives a label from 5 participants, enabling us to compare multiple users' perceptions of the same action plan. The full survey is included in Appendix~\ref{appendix-survey}.

\subsubsection{Participants}
\rthree{\label{sec:user-survey-participants} We conducted our survey with IRB oversight. We recruited $N=20$ participants} \aeone{via email, snowball, and word-of-mouth. Our aim is for a relatively uniform distribution of participants across age groups since age is a key factor in smart home perceptions and satisfaction~\cite{shin2018will}. Our resulting participant pool is semi-uniformly distributed with $20\% \pm 5\%$ in each of 6 age groups, with only the 55-64 group underrepresented at $5\%$. 
We compensated participants with \$20 gift cards.}

\begin{table}
    \centering
    \begin{tabular}{llllll}
    \hline
       \textbf{ID} & \textbf{Gender}   & \textbf{Age}   & \textbf{Education}                      & \textbf{Smart Home Familiarity}   & \textbf{English Level}    \\
    \hline
 SP1   & F        & 45-54 & Bachelor's                     & Moderately              & Native/bilingual \\
 SP2   & M        & 65+   & Graduate                       & Very                    & Native/bilingual \\
 SP3   & F        & 55-64 & Bachelor's                     & Slightly                & Native/bilingual \\
 SP4   & F        & 25-34 & Bachelor's                     & Slightly                & Native/bilingual \\
 SP5   & M        & 25-34 & Graduate                       & Moderately              & Native/bilingual \\
 SP6   & F        & 18-24 & High school or equivalent      & Moderately              & Native/bilingual \\
 SP7   & M        & 18-24 & High school or equivalent      & Moderately              & Native/bilingual \\
 SP8   & M        & 25-34 & Graduate                       & Slightly                & Native/bilingual \\
 SP9   & F        & 65+   & Some college                   & Moderately              & Intermediate     \\
 SP10  & M        & 65+   & Associates or technical degree & Moderately              & Native/bilingual \\
 SP11  & F        & 25-34 & Graduate                       & Moderately              & Native/bilingual \\
 SP12  & M        & 65+   & Some college                   & Moderately              & Native/bilingual \\
 SP13  & F        & 18-24 & Associates or technical degree & Very                    & Native/bilingual \\
 SP14  & F        & 25-34 & Graduate                       & Moderately              & Native/bilingual \\
 SP15  & F        & 35-44 & Graduate                       & Moderately              & Native/bilingual \\
 SP16  & F        & 18-24 & High school or equivalent      & Very                    & Native/bilingual \\
 SP17  & F        & 45-54 & Graduate                       & Slightly                & Native/bilingual \\
 SP18  & M        & 35-44 & Some college                   & Slightly                & Basic            \\
 SP19  & M        & 45-54 & Bachelor's                     & Moderately              & Native/bilingual \\
 SP20  & F        & 35-44 & Bachelor's                     & Very                    & Native/bilingual \\
    \hline
    \end{tabular}
    \caption{\rthree{Demographic characteristics of survey participants (SP) in our empirical study of LLM-based smart home control.}}
    \label{tab:participants}
\end{table}

\subsubsection{Analysis: User-perceived quality} We score the user-perceived quality of an action plan based on the satisfaction rating assigned from our user survey. We map each categorical rating given on the five-point scale from ``Very unsatisfied'' to ``Very satisified'' to a numerical rating from $[0.2, 1.0]$. For further qualitative analysis, we study free-form rationales on action plans that have above- or below-average ratings, or have high variability among their five labelers.

\subsubsection{Analysis: Failure modes} \aeone{We derive failure modes from the free-form rationales in our user survey using a systematic two-phase coding process:

\begin{enumerate}
    \item One researcher reviewed all responses to identify several high-level categories of failure. Three researchers then independently assigned labels from these high-level categories to the 325 responses with less than ``Very satisfied'' quality.
    \item The same three researchers discussed more specific modes of failure they observed during the high-level categorization task. After defining a set of finer-grained failure modes, each researcher independently assigned these labels to each of their original annotations.
\end{enumerate}

This process yielded a total of $975$ annotations with a Krippendorff's alpha of $0.75$ and pairwise Cohen's kappa of $0.75, 0.76, $ and $0.74$, suggesting acceptable inter-annotator agreement.}

\aetwo{\subsection{Action Plan Analysis Procedures \& Metrics}
\label{sec:action-plan-procedures}}
\subsubsection{Analysis: Relevance metrics} \label{sec:relevance-metrics} \aeone{We analyze action plans by examining how relevant the target devices are to the goal category of the command. We loosely base these metrics on \emph{document relevance} in information retrieval~\cite{cooper1971definition}. For each of the 7 goal categories, we identify, as a sort of ground truth, the types of device that are relevant (based on the types of device targeted by IFTTT routines in that goal category~\cite{yu2021analysis}). Given our set of user commands $C$, the goal category $g_c$ of command $c \in C$, a set of the goal categories of devices $G_c$ that an LLM's action plan targets in response to command $c$, and the set of goal categories $G_h$ supported by devices in home $h$, we measure the following:

\textbf{False positives ($FP$):} a false positive occurs when the system creates a plan that targets one or more devices, but the home in fact does not have any devices relevant to the command. We count these as:
\setlength{\columnsep}{-3cm}
\vspace*{-2.25\multicolsep}
    \begin{multicols}{2}
        \begin{equation} \nonumber
            FP = \frac{1}{|C|}\sum_{c \in C}^{}{{\it fp}(G_c, G_h, g_c)} 
        \end{equation}\break
        \begin{equation} \nonumber
            {\it fp}(G_c, G_h, g_c) = 
            \begin{cases}
              1 & \text{if $|G_c| > 0$ and $g_c \notin G_h$} \\
              0 & \text{otherwise}
            \end{cases} 
        \end{equation}
    \end{multicols}
\textbf{False negatives ($FN$):} in contrast, a false negative occurs when the system does not create a plan, but the home has relevant devices:
\vspace*{-2.25\multicolsep}
    \begin{multicols}{2}
        \begin{equation} \nonumber
            FN = \frac{1}{|C|}\sum_{c \in C}^{}{{\it fn}(G_c, G_h, g_c)} 
        \end{equation}\break
        \begin{equation} \nonumber
            {\it fn}(G_c, G_h, g_c) = 
            \begin{cases}
              1 & \text{if $|G_c| = 0$ and $g_c \in G_h$} \\
              0 & \text{otherwise}
            \end{cases} 
        \end{equation}
    \end{multicols}
    
\textbf{Accuracy ($Acc$):} the accuracy captures the portion of plans that target a relevant device only when one exists: 
$$Acc = 1 - (FP + FN)$$

\textbf{Relevance score ($Rel$):} finally, we measure the relevance of system outputs by computing, for each command in $C$, a score $r \in [-1, 1]$, assigned based on the relative number of relevant and irrelevant devices included in each command's generated action plan. Lower numbers indicate more of the target devices are irrelevant, while higher numbers indicate more are relevant. We define $Rel$ as the average across all $r$ for a system's outputs, as follows:

\vspace*{-2.25\multicolsep}
\begin{multicols}{2}
    \begin{equation} \nonumber
        Rel = \frac{1}{|C|}\sum_{c \in C}^{}{r(G_c, G_h)}
    \end{equation}\break
    \begin{equation} \nonumber
        {\it r}(G_c, G_h) = \frac{|G_c \cap G_h| - |G_c - G_h|}{|G_c|}
    \end{equation}
\end{multicols}
    
When comparing performance between LLMs, we scale $Rel$ based on the portion of action plans that do not reject the command, since LLMs that reject higher numbers of commands would otherwise have inflated $Rel$.}

\subsubsection{Analysis: Cost and latency} The number of tokens that are input to or output from the LLM are an indicator of cost ~\cite{webster1992tokenization}. 
We measure tokens based on the encoding scheme used by a given model~\cite{kudo2018sentencepiece, openai2023}. We also measure response latency (in seconds) by timing the function in our code that calls the relevant remote API. For GPT models, we use the OpenAI API; for Llama, we use a dedicated deployment on a Hugging Face ``Inference Endpoint''. In both cases, we access the API from a 1Gbps fiber internet connection. Latency includes network delays, varying loads on distributed computing resources, etc., and provides an estimate of user-facing performance.

\subsubsection{Analysis: JSON validity} We note if 
forced extraction of JSON was necessary and possible. First, we attempt to decode the response verbatim---in this case, no extraction is necessary. If this fails, we attempt to extract and decode the portion of the response enclosed in top-level curly brackets. This enables us to identify the valid JSON portion of responses 
that include additional exposition (e.g., ``Sure, here is the JSON action plan: \texttt{\{ ... \}}''). Finally, if this also fails, we attempt to interpret the extracted string as an abstract syntax tree, enabling us to decode the response even if it contains, e.g., portions that are expressed in Python syntax rather than JSON. If all steps fail, we consider the response invalid.

\subsubsection{Correlation testing} \aeone{We use Kendall's tau-b ($\tau_b$) to test for correlations between quality labels in our user survey and characteristics of action plans. $\tau_b$ tests for correlation between two variables that may not obey normal distributions, with $-1$ and $1$ implying strong negative and positive correlations, respectively, and $0$ implying no correlation. Given the degree of subjectivity in our task (particularly with respect to user-perceived quality ratings), we set the threshold for $\tau_b \geq 0.2$ or $\tau_b \leq -0.2$ with $p \leq 0.05$. Our threshold for $\tau_b$ obeys a common guideline for establishing at least ``moderate'' correlation~\cite{botsch2011chapter}.}

\aetwo{\subsection{Findings}
\label{sec:findings}}
\subsubsection{Finding 1: Action plans vary from remarkably helpful to completely non-sensical} \label{sec:empirical-finding-1} Toward answering \textbf{RQ1-2}, we 
highlight several action plans based on indicators in their user-perceived quality ratings. Table~\ref{tab:plans-and-rationales} depicts action plans with (1)~consistently high, (2)~consistently low, and (3)~highly-variable quality ratings from our survey. 
Consistently high- or low-quality plans are rated higher or lower quality on average than the population, with below-average variance. Highly-variable plans have above-average variance in their quality ratings.

\bgroup
\def\arraystretch{1.15}
\begin{table}
    \centering
\begin{tabular}{l|p{12.5cm}} \hline
\multicolumn{2}{l}{\textbf{High Satisfaction} (47\% of plans)} \\ \hline
\multicolumn{2}{l}{\cellcolor{green!5} $h_1$ help me lower my power bill} \\
\multicolumn{2}{l}{\cellcolor{green!5} \parbox{15cm}{To lower your power bill, I will turn off the lights in any room where there has been no motion detected for 15 minutes. This will ensure that lights are not left on unnecessarily.}} \\ \hline

Very satisfied & SP1: Good choice. \\ 
Very satisfied & \parbox{12.5cm}{SP7: Very nice answer, as well as understanding and figuring out the required action to achieve the task. Since there are only lights, home fully understood how to save energy.} \\
Very satisfied & SP16: Yeah that's great \\ \hline 

\multicolumn{2}{l}{\cellcolor{green!5} $h_2$ help me wind down} \\
\multicolumn{2}{l}{\cellcolor{green!5} \parbox{15cm}{I have dimmed the lights in the living room and bedroom to a warm color, turned off the lights in other rooms, set the thermostat to a comfortable 68 degrees, and lowered the volume of the smart speaker. The television is turned off to help you wind down.}} \\ \hline

Very satisfied & SP10: Good strategy with devices available. \\ 
Very satisfied & SP16: This is solid. The AI has a better idea of how to wind down than I do. \\
Satisfied & SP17: This sounds lovely for winding down. \\ \hline \hline

\multicolumn{2}{l}{\textbf{Low Satisfaction} (13\% of plans)} \\ \hline

\multicolumn{2}{l}{\cellcolor{red!5} $h_2$ stop cleaning after sunset} \\
\multicolumn{2}{l}{\cellcolor{red!5} \parbox{15cm}{When the local time sensor detects that it is sunset, the smart speaker in the living room will be commanded to stop the cleaning process.}} \\ \hline

Very unsatisfied & SP1: The speaker doesn't clean. \\
Unsatisfied & SP8: Smart speaker seems irrelevant. \\
Very unsatisfied & SP18: Speakers do not clean. \\ \hline

\multicolumn{2}{l}{\cellcolor{red!5} $h_1$ don't clean when people are here} \\
\multicolumn{2}{l}{\cellcolor{red!5} \parbox{15cm}{When motion is detected in the entry, the cleaning robot will be turned off.}} \\ \hline

Very unsatisfied & \parbox{12.5cm}{SP7: There is no cleaning robot in this house. Plus, checking for the motion in the entry will not tell you if there are people in the house or not.} \\

Neutral & SP12: There is no robot. \\

Very unsatisfied & SP18: No cleaning robot in house. \\ \hline \hline

\multicolumn{2}{l}{\textbf{Variable Satisfaction} (40\% of plans)} \\ \hline

\multicolumn{2}{l}{\cellcolor{yellow!5} $h_1$ let the guest in} \\
\multicolumn{2}{l}{\cellcolor{yellow!5} \parbox{15cm}{The entry overhead light has been turned on with 60\% brightness to let the guest in.}} \\ \hline

Satisfied & \parbox{12.5cm}{SP6: The most useful device available was used, so seems like a good response.} \\

Very unsatisfied & \parbox{12.5cm}{SP7: There is no device that controls the entrance door, thus the home should just say that it can't do this task since there is no device it could use.} \\

Neutral & SP19: Turning on a light helps the guest enter, but that's different from letting them in. \\ \hline

\multicolumn{2}{l}{\cellcolor{yellow!5} $h_2$ turn off the thermostat when I don't need it} \\
\multicolumn{2}{l}{\cellcolor{yellow!5} \parbox{15cm}{The thermostat will be turned off when the temperature sensor in the living room indicates that the temperature is within a comfortable range, suggesting that the thermostat is not needed.}} \\ \hline

Unsatisfied & SP1: I might not need it based on the fact that I am not home. \\

Neutral & SP4: I would also want the home to stop heating or cooling when I’m absent. \\

Very satisfied & SP18: The home grasped concept of “need it” \\ \hline

\end{tabular}
    \caption{GPT-4 generated action plans with consistently high, consistently low, and variable user satisfaction across labelers. Consistent high or low satisfaction implies the plan is either helpful or exhibits obvious failures, respectively, with low subjectivity. Variable satisfaction implies that some aspects of the plan are subject to individual preferences.}
    \label{tab:plans-and-rationales}
\end{table}
\egroup

\begin{figure}
    \centering
    \includegraphics[width=0.6\textwidth]{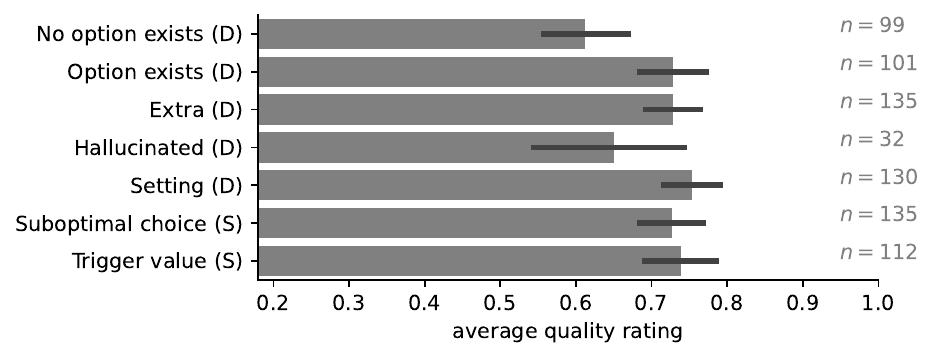}
    \caption{User-rated quality of GPT-4 generated action plans by failure mode across all homes. (D) denotes device targeting failures; (S) denotes sensor choice failures. Error bars depict a 95\% confidence interval. $n$ indicates the number of samples (i.e., labels assigned to action plans) for each failure mode. Users rate \emph{false positives} more harshly than other failure modes: ``No option exists'' and ``Hallucinated'' have the lowest average quality.}
    \label{fig:failure-mode-quality}
\end{figure}

\changed{\textbf{High quality plans provide complex solutions to completely under-specified commands.} In the case of ``help me lower my power bill'', GPT-4 relates the only device available in home $h_1$ (lights) to the under-specified goal of lowering a power bill. It also proposes a sensor trigger and value (motion for 15 minutes) that will accomplish it. In the case of ``help me wind down'', GPT-4 relates the activity of ``winding down'' to an extensive set of devices and settings to accomplish the goal. Users rate these plans high quality. 

\textbf{Low quality plans exhibit failures that diminish user satisfaction and limit practical utility.} In the case of ``stop cleaning after sunset'' in home $h_2$ (which has no cleaning capability), GPT-4 targets the speaker with a non-sensical plan; in the case of ``don't clean when people are here'' in $h_2$, GPT-4 chooses a suboptimal trigger and plans to target a cleaning robot that does not exist in the home. Users rate these plans low quality.

\textbf{Plans with variable quality involve ``creative,'' subjective choices.} This is exemplified by the plan to ``let the guest in'' in home $h_1$. One user finds GPT-4's decision to turn on the entryway light as a reasonable response given the home's limitations, while another finds it non-sensical and prefers that the system reject the request. In the case of ``turn off the thermostat when I don't need it'', users' definitions of ``don't need it'' vary, resulting in differing levels of satisfaction. Whether these plans are seen as resourceful (and thus high quality) or non-sensical (and thus low quality) varies based on user preference.

\subsubsection{Finding 2: Action plans exhibit specific modes of failure that diminish user satisfaction} 
\label{sec:empirical-finding-2}
We identify 7 common failure modes in LLM-generated action plans, as follows:

\begin{enumerate}
    \item \textbf{Device $\rightarrow$ No option exists:} The action plan targets an irrelevant device and no better option (other than refusing the request) appears to exist, e.g., turning off the smart speaker in response to ``stop cleaning'' when there is no cleaning robot in the house.

    \item \textbf{Device $\rightarrow$ Option exists:} The action plan does not target a relevant device even when an option exists, either because the request is refused or the plan fails to include additional relevant targets, e.g., adjusting the lights but not the thermostat in order to ``save energy''.

    \item \textbf{Device $\rightarrow$ Extra:} The action plan targets an extra, irrelevant or relevant device or devices in addition to relevant devices, e.g., turning on both the TV and the smart speaker.

    \item \textbf{Device $\rightarrow$ Hallucinated:} The action plan targets a device that does not exist in the home, i.e., the LLM ``hallucinates'' a device and adds it to the action plan, e.g., turning on the cleaning robot when there is no cleaning robot in the home.

    \item \textbf{Device $\rightarrow$ Setting:} The action plan assigns a setting that may be incorrect or dependent on additional context, e.g., setting the lights to full brightness regardless of the time of day.

    \item \textbf{Sensor $\rightarrow$ Suboptimal choice:} For persistent goals, the action plan chooses a suboptimal sensor for the automation, e.g., using motion in the entryway to determine occupancy.

    \item \textbf{Sensor $\rightarrow$ Trigger value:} For persistent goals, the action plan chooses an illogical or under-specified trigger value for the automation, e.g., using the time value of ``sunset'' as a trigger to turn off the lights.
\end{enumerate}

We illustrate an action plan and corresponding rationale that exemplifies each failure mode in Appendix~\ref{appendix-failure-modes}. 

\textbf{False positives are judged most harshly by users.} Fig.~\ref{fig:failure-mode-quality} depicts the average user-perceived quality of action plans that exhibit each failure mode. ``No option exists'' has the lowest average quality, followed by ``Hallucinated''. This implies that \emph{false positives}---either due to targeting of an extant but irrelevant device, or due to an attempt to target a device that does not exist---are judged harshly by users. We test for correlation between the average quality of an action plan and its false positive rating ${\it fp}$, with $\tau_b = -0.42$ and $p = 0.00000004$, suggesting a strong negative correlation with high significance.}

\changed{\textbf{Failure modes are modulated by the capabilities of the home}. Fig.~\ref{fig:failure-modes} depicts the frequency of each failure mode, grouped by home. Accounting for all commands, when moving from home $h_1$ to $h_3$ the frequency of ``No option'' and ``Hallucinated'' failures decreases since increasingly more devices are available to meet the goal. Inversely, the frequency of ``Option exists'' failures \emph{increases}: when more devices are available, users expect them to be targeted. Accounting only for commands with goals that are supported by the home, the dominant failure mode is ``Extra''---action plans tend to target extra devices in cases where the goal is achievable. 

\begin{figure}[t!]
    \centering
    \includegraphics[width=0.49\columnwidth]{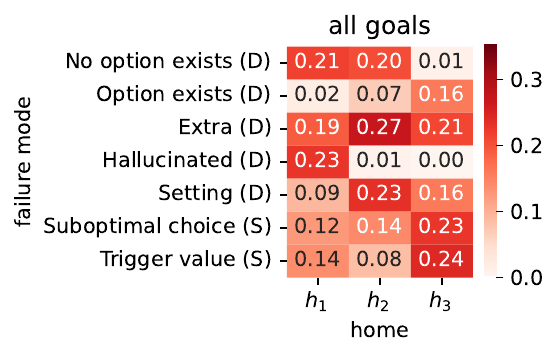} 
    \hfill
    \includegraphics[width=0.49\columnwidth]{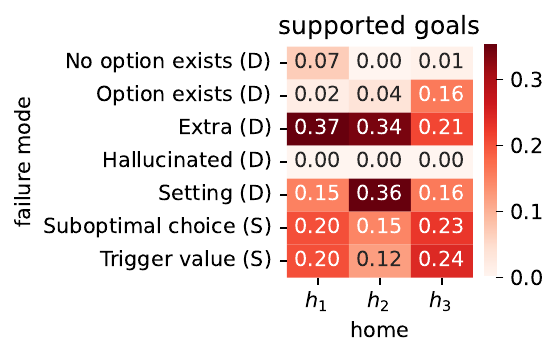}
    \caption{Frequency of failure modes in GPT-4 generated action plans, organized by home ($h_1, h_2$, and $h_3$, each with an increasing number and diversity of devices). Left: action plans for all commands. Right: action plans for commands with goals that are supported by the home (e.g., the plan for ``help me cool off'' in home $h_1$ is omitted since $h_1$ does not have a thermostat). (D) denotes device targeting failures; (S) denotes sensor choice failures.}
    \label{fig:failure-modes}
\end{figure}

\subsubsection{Finding 3: Reliably targeting relevant devices is a challenge, even for the highest-performing models.}\label{sec:finding3}
\begin{table}
\centering
\begin{tabular}{llr||c|c|c|c||r|r|r||r|r}
\hline
\multicolumn{3}{l}{\textbf{Configuration}} & \multicolumn{4}{c}{\textbf{Relevance}} & \multicolumn{3}{c}{\textbf{Latency (s)}} & \multicolumn{2}{c}{\textbf{JSON}} \\ \hline
\multicolumn{2}{l}{Model} & \textbf{$t$} & \textbf{$Acc$} & \textbf{$FP$} & \textbf{$FN$} & \textbf{$Rel$} & Min & Max & Mean & Forced & Invalid \\ \hline

\hline
\rowcolor{cyan!10}
 GPT-4   &                  &  0.00 &   0.84 & 0.16 & 0.00 &  0.63 &          0.82 &         35.83 &           9.54 &          0.01 &           0.00 \\
 GPT-4   &                  &  0.70 &   0.82 & 0.16 & 0.02 &  0.67 &          0.71 &         46.69 &          12.59 &          0.01 &           0.00 \\
 GPT-3.5 & text-davinci-003 &  0.00 &   0.68 & 0.32 & 0.00 &  0.30 &          0.47 &         20.58 &           6.91 &          0.43 &           0.00 \\
 GPT-3.5 & text-davinci-003 &  0.70 &   0.68 & 0.33 & 0.00 &  0.26 &          0.51 &         23.75 &           7.46 &          0.37 &           0.00 \\
 Llama 2 & 7B-chat-hf       &  0.10 &   0.65 & 0.28 & 0.07 &  0.09 &          1.23 &         26.12 &          12.79 &          0.61 &           0.38 \\
 Llama 2 & 7B-chat-hf       &  0.70 &   0.62 & 0.28 & 0.10 &  0.11 &          1.17 &         16.61 &           9.94 &          0.59 &           0.38 \\
 Llama 2 & 13B-chat-hf      &  0.10 &   0.62 & 0.28 & 0.09 &  0.18 &          3.17 &         49.87 &          30.05 &          0.79 &           0.19 \\
 Llama 2 & 13B-chat-hf      &  0.70 &   0.62 & 0.29 & 0.09 &  0.21 &          2.97 &         58.05 &          29.93 &          0.77 &           0.22 \\
\hline
\end{tabular}

\caption{Summary of quantitative results from our empirical study. We report relevance metrics, which reflect an LLM's tendency to include devices that are irrelevant to the goal in an action plan (accuracy $Acc$, false positives $FP$, false negatives $FN$, and relevance score $Rel$). We also report response latency and JSON validity. Temperature $t$ sets the probability that the LLM predicts less-likely tokens in the output sequence (i.e., the ``creativity'' of output). The shaded row denotes the set of action plans that participants rated in our user survey.}
\label{tab:quantitative-results}
\end{table}
We report accuracy, false positives, false negatives, and relevance scores for 4 different models, each with 2 different temperature $t$ parameters in Table~\ref{tab:quantitative-results}. $t$ sets the probability that the LLM predicts less-likely tokens in the output sequence (i.e., the ``creativity'' of output).

\begin{figure}[t!]
    \centering
    \includegraphics[width=0.49\columnwidth]{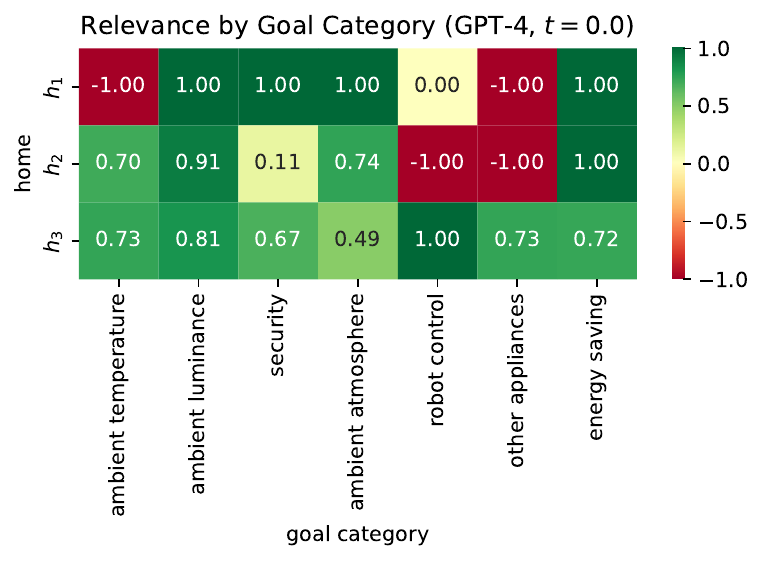} 
    \hfill
    \includegraphics[width=0.49\columnwidth]{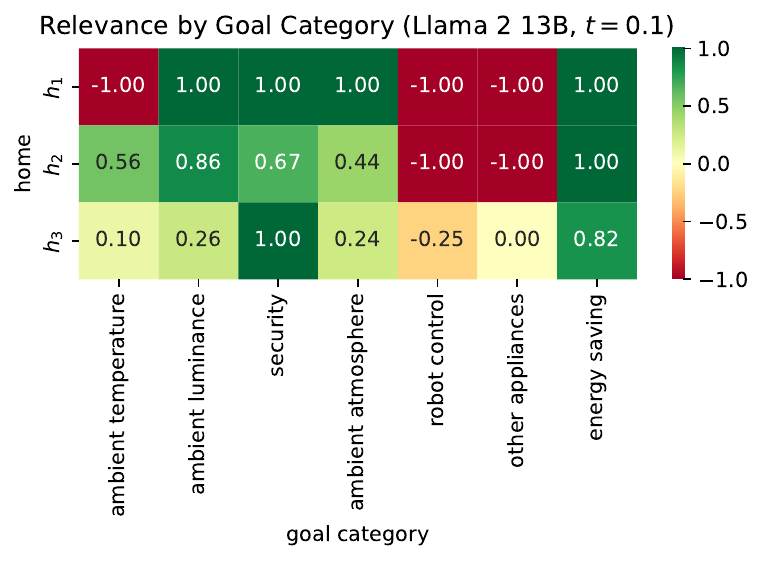}
    \caption{Relevance scores of LLM-generated action plans, grouped by home ($h_1, h_2,$ and $h_3$, which have an increasing number and diversity of devices) and the goal category of commands (which determines the types of devices that are relevant). -1.0 implies all devices in all action plans for commands in that category are irrelevant; 0.0 implies either no target, or an equal balance of relevant to irrelevant; 1.0 implies all are relevant. Action plans in homes with fewer devices ($h_1$) are more likely to target a minimal relevant set (1.0), but also target completely irrelevant devices when the goal is not supported (-1.0). In homes with more capabilities ($h_3$), action plans include both relevant and irrelevant devices, diluting the relevance score.}
    \label{fig:relevance-by-category}
\end{figure}

GPT-4 outperforms other models on all metrics. All models' relevance scores are far from 1.0. GPT-4 $t=0.7$ achieves the highest at $0.63$, indicating that GPT-4 is the most likely to target relevant devices and not target irrelevant ones. $0.63$ nonetheless leaves much room for improvement. All models tend to target extra, irrelevant devices in addition to the minimal relevant set as the home provides more capabilities. We depict this in Fig.~\ref{fig:relevance-by-category} by grouping action plan relevance scores by homes and goal categories of commands. In homes with fewer capabilities (e.g., $h_1$), action plans tend to target the minimal relevant set of devices (achieving 1.0 in several goal categories), but also target completely irrelevant sets of devices with non-sensical plans when the goal is unsupported (achieving -1.0 in these categories). As the home gains capabilities, the risk of creating non-sensical plans decreases (shown by fewer -1.0 scores in $h_2$ and $h_3$). This improvement, however, is moderated by the fact that action plans begin to include extra, irrelevant devices in addition to the minimal relevant set.}

\subsubsection{Finding 4: Cost and latency pose practical challenges}
\textbf{Complex homes and goals incur higher costs.} For immediate goals (which do not require sensors to be included in the template), input tokens are 469, 529, and 607 for $h_1, h_2,$ and $h_3$, respectively; for persistent goals, these values increase to 670, 730, and 808. Overall, output tokens vary from 8 to 566 depending on the goal. In the minimal case, the output is simply to reject the goal. In the maximal case, the output is a lengthy action plan. Based on these token measurements in conjunction with API pricing, a single smart home command costs around \$0.02 for GPT-3.5 and \$0.03 for GPT-4.\footnote{OpenAI API pricing in November 2023: \$0.02 per 1K tokens for GPT-3.5 and \$0.03 per 1K input tokens	/ \$0.06 per 1K output tokens for GPT-4} Accounting for longitudinal data about how often users interact with home assistants~\cite{bentley2018understanding}, this would range from \$0.10 to \$0.60 per day for GPT-3.5, or \$0.30 to \$1.00 per day for GPT-4.

\textbf{High performance models and complex goals incur longer response times.} Latency statistics are shown in Table~\ref{tab:quantitative-results}. Response latency varies widely, dependent on external factors (e.g., demand on cloud infrastructure) and the complexity of the response needed. In the best case, latency for all models is around one second; this occurs when the request is rejected. Latency for generating more complex action plans, however, is highly unstable, often approaching 30 to 45 seconds for the most powerful models.

\subsubsection{Finding 5: JSON response validity poses a practical challenge} Table~\ref{tab:quantitative-results} depicts the portion of action plans that required either forced extraction or were completely invalid. Nearly all models' outputs occasionally need forced extraction. This is frequent with the Llama 2 models, which have been fine-tuned for chat applications. Responses often included a valid action plan preceded by exposition: e.g., ``Sure, here is the action plan...'' Across GPT models, there was no invalid JSON; Llama 2, however, occasionally responded without JSON, but rather a description of the plan, or the JSON separated into several segments. This is an engineering challenge: fine-tuning increases confidence that a model's outputs will have a desired structure~\cite{wei2021finetuned}. 

\subsubsection{Summary} \label{sec:empirical-motivation-findings} We draw the following conclusions from our empirical study:
\begin{itemize}
    \item LLMs can generate creative and complex action plans in response to under-specified user commands. \textbf{\hyperref[rq1]{(RQ1)}}
    \item LLMs are prone to failures surrounding the relevance of target devices, their assigned settings, and in the case of persistent goals, sensor triggers and trigger values. False positives (i.e., targeting an irrelevant or nonexistent device when none are available) are negatively correlated with user satisfaction. \textbf{\hyperref[rq2]{(RQ2)}}
    \item Performance is compared in terms of accuracy, false positives, false negatives, and relevance scoring of models' action plans. Even the highest-performing models struggle to reliably target relevant devices: they either (1) target irrelevant devices when the home has no devices to support the goal or (2) target extra, irrelevant devices in addition to the minimal relevant set when the home has more devices available. \textbf{\hyperref[rq2]{(RQ2)}}
    \item The cost and latency of generating action plans, in addition to instability in the validity of the output are open engineering challenges. \textbf{\hyperref[rq2]{(RQ2)}}
\end{itemize}

\section{Designing Sasha}
\label{approach}
\begin{figure}[t!]
    \centering
    \vspace{.1cm}
    \includegraphics[width=\columnwidth]{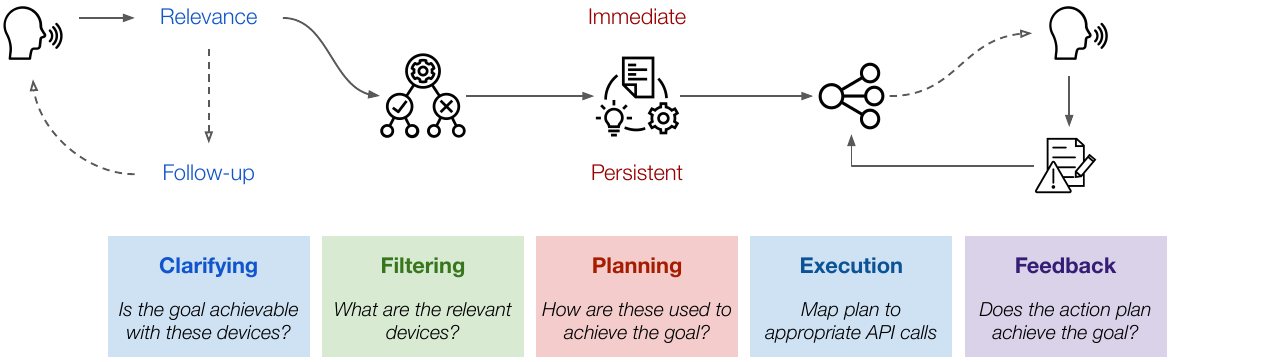}
    \caption{Sasha uses iterative reasoning to generate consistently high-quality action plans that leverage whichever relevant devices are available in a given smart home.}
    \label{fig:sasha}
\end{figure}

The prototype LLM-based smart home presented in Fig.~\ref{fig:prototype} and used in our empirical study above makes some remarkable strides towards addressing \textbf{RQ3.} However, as our empirical study shows, there are 
shortcomings of the na\"{i}ve approach that must be addressed. In this section, we propose Sasha, a \underline{s}m\underline{a}rter \underline{s}mart \underline{h}ome \underline{a}ssistant that improves the performance of LLMs in smart environments. \rfour{We are motivated by} \aeone{the need to improve response quality by (1)~reducing false positives, (2)~increasing the relevance of action plans, and (3)~providing input for user feedback to reduce subjectivity. Our empirical study revealed each of these as key factors in user satisfaction.} 

Sasha deconstructs the process of responding to user commands into \emph{iterative reasoning components} (Fig.~\ref{fig:sasha}). Rather than creating an action plan in a single inference, Sasha reasons about the response in multiple steps. We assume the use of an \emph{instruction-tuned} model that has been fine-tuned to adhere to written instructions provided in prompts~\cite{wei2021finetuned} but is not fine-tuned for smart home tasks. Furthermore, we assume that the model exhibits high performance with \emph{chain of thought prompting}~\cite{wei2022chain}. Both are true of numerous models~\cite{chowdhery2022palm, brown2020language, thoppilan2022lamda, openai2023gpt4}. The two inputs to Sasha remain the same as in Section~\ref{feasibility-study}: a user command and a template that describes the sensors and devices available. We describe each component of Sasha's iterative reasoning in the following.

\subsection{Clarifying: Is the goal achievable with the devices available?}
\label{sec:sasha-clarifying}
\emph{False positives} are negatively correlated with user satisfaction, as shown in our empirical study. Our first objective is to develop a component that directly reduces the rate of false positives. The \emph{clarifying} step therefore prompts the model to consider the goal of the user's command in relation to the devices available in the home template. \rfour{If the home has devices that are semantically relevant to the command, Sasha will determine that the goal is achievable and move beyond the clarifying step without interjection from the user.} If Sasha does not identify a sufficient set of relevant devices, it notifies the user that it is unable to proceed. We allow the user to follow-up with clarifying information, at which point the prompt is repeated with added context. The output of this component is a status code: success if there are relevant devices, and failure if not.

We illustrate with an example. When a user tells Sasha ``I'm tired'', we prompt the model with the command, the devices in the home, and instructions to consider whether any of the devices are relevant to the command. If the command is issued in, e.g., home $h_3$ from our empirical study (with a coffee pot), Sasha will relate ``tired'' to the coffee pot and determine that there do exist relevant devices in the home. If, instead, the same command is issued to the same system in a home with only lights ($h_1$), Sasha is unlikely to find devices that seem relevant. The output is ``failure'', along with a verbal response to the user: ``I'm sorry, I didn't find anything in the home that seems relevant to your command. Can you provide more information?'' For the sake of example, we assume the user is tired and the lights will help them wake up. The user can provide clarifying information (again in their own phrasing): ``I need help waking up''. We repeat the prompt with the added context, asking the system to again consider whether any devices are relevant. In $h_1$ with only lights, the model is likely to find semantic relationships between ``tired'', ``waking up'', and the color and brightness of lights, so it will now return ``success''.

It is also important to consider cases where the home simply cannot support the goal, which we revealed in our empirical study can produce non-sensical output. Let us assume that the user follows up ``I'm tired'' with clarification: ``I could really use some coffee''. At this point, there is sufficient information to infer a goal, but in the case of $h_1$ and $h_2$ there are still no relevant devices to support it. Sasha in this case will return ``failure'' for relevance and avoid producing a non-sensical action plan. Without Sasha's relevance check, the model would turn on a light in $h_1$ and justify its reasoning: ``Turning on the entry overhead light to provide a bright, energizing atmosphere to help you get ready for your coffee.''

Performing the clarifying step of reasoning with an LLM provides several advantages over a task-specific component, as task-specific approaches have two key limitations: 
(1)~their design encodes a constrained set of assumptions about which devices are pertinent to which goals, which precludes the invention of \emph{creative} action plans, and (2)~there exist generic device types (e.g., a smart plug) for which it is nearly impossible to predict their relationship to a user's goal. Using an LLM provides an advantage in both cases: (1)~it can leverage the often-unpredictable situational knowledge encoded in its training corpora to invent creative action plans, and (2)~it reasons \emph{semantically} about the devices. This means that simply \emph{naming} a smart plug in the home template, e.g., \texttt{humidifier\_plug} or \texttt{guitar\_amp\_plug}, 
enables a model to leverage the same device type for a broader set of actions than those assumed 
in task-specific systems. As an implementation note, we have the option to perform this reasoning on either \emph{the full device template} (i.e., a semantically complete and correct JSON snippet) or, instead, on a shortened \emph{list of available devices} (i.e., simply a list of the home-provided names of the devices). The latter option benefits performance, which we highlight in our evaluation.

\subsection{Filtering: Which are the relevant devices?}
\label{sec:sasha-filtering}
\aeone{\emph{Relevance} of target devices in action plans tends to be unstable across homes, as shown in our empirical study. Our next objective is to develop a component that provides consistent relevance ($Rel$) across different environments.}
To do so, we prompt the model to select the minimal set of relevant devices from the JSON home template. This step outputs a JSON subset of the devices in the home template.

For the ``I'm tired, I need help waking up'' command, we prompt the model with the device filtering prompt. In response, the model in $h_3$ will return JSON of the coffee maker in the kitchen. 
In $h_1$ or $h_2$, it will return JSON of lights in their respective rooms. \aeone{As with the clarifying step, we may either filter the full set of devices, or filter a list of devices (with their settings excluded).} Performing this step with an LLM as opposed to a task-specific system has a similar justification as above. Filtering based on the semantic relationship between a command and the named devices in a home unlocks greater flexibility than a system that relies on a rigidly-defined ontology.

\subsection{Planning: How are these devices and sensors used to meet the goal?}
After the model has selected the relevant devices, we prompt it to produce an action plan. The planning prompt includes the user command, the filtered devices (and, for persistent goals, the sensors). In the case of immediate goals, we guide the model to produce an action plan that is simply a modified version of the filtered subset of the home template, with settings assigned to devices. For persistent goals, we guide the model to produce an action plan in the form of an automation routine, defined as a (trigger, action) pair. The trigger is a sensor or set of sensors and corresponding values that will trigger the routine, while the action is the filtered subset of devices in the home template with settings assigned. This step outputs an action plan, in JSON.

\subsection{Execution: Map the plan to API calls}
We execute action plans by mapping changes to device state in the JSON to the appropriate smart home API calls. This is a matter of implementation and requires no help from the model.

\subsection{Feedback: Does the action plan achieve the user's goal?}
\label{sec:sasha-feedback}
\aeone{The user-perceived quality of an action plan is sometimes subjective, as shown in our empirical study. Our last objective is to provide input for users to occasionally refine aspects of generated action plans.}
After execution, the model's action plan is described to the user for them to critique. \rfour{If the user is satisfied with the choices made in the action plan, no further input is needed.} \rfour{If some aspects of the plan need adjustment, however, the user has an opportunity to provide feedback.} They can provide this feedback in unconstrained language (as with the initial command). For example, if the user tells Sasha in $h_3$ ``I'm bored'' and the system responds by turning on the TV and smart speaker, the user may offer some feedback: ``I don't need both the TV and the speaker to be on''. Once the user provides this feedback, we prompt the model to modify the plan, or apologize and take no action if Sasha finds no way to improve it. This step outputs verbal responses to the user describing the action plan as it is progressively modified, along with the final action plan JSON. 

\section{Evaluation}
\label{evaluation}
\begin{table}
\centering
\begin{tabular}{ll|c||c|c|c|c||rrr||rr}
\hline
\multicolumn{3}{l}{\textbf{Configuration}} & \multicolumn{4}{c}{\textbf{Relevance}} & \multicolumn{3}{c}{\textbf{Latency (s)}} & \multicolumn{2}{c}{\textbf{JSON}} \\ \hline
\multicolumn{2}{l}{Model} & List & \textbf{$\Delta Acc$} & \textbf{$\Delta FP$} & \textbf{$\Delta FN$} & \textbf{$\Delta Rel$} & Min & Max & Mean & Forced & Invalid \\ \hline

\multicolumn{9}{l}{\emph{Clarifying + Filtering}} \\ \hline
 GPT-4   & gpt-4            &   & \cellcolor{green!10}+0.10 & \cellcolor{green!13}-0.12 & \cellcolor{red!3}+0.03 &       &          0.72 &         44.54 &          11.90 &          0.00 &           0.00 \\
 GPT-4   & gpt-4            & \checkmark & \cellcolor{green!7}+0.07  & \cellcolor{green!9}-0.08  & \cellcolor{red!2}+0.02 &       &          0.71 &         14.43 &           4.84 &          0.00 &           0.00 \\
 GPT-3.5 & text-davinci-003 &  & \cellcolor{green!1}+0.01  & \cellcolor{green!1}-0.01  & 0.0                    &       &          0.40 &         19.45 &           6.26 &          0.00 &           0.00 \\
 GPT-3.5 & text-davinci-003 & \checkmark & \cellcolor{green!2}+0.02  & \cellcolor{green!2}-0.02  & 0.0                    &       &          0.40 &          6.06 &           2.80 &          0.01 &           0.00 \\
\hline

\multicolumn{9}{l}{\emph{Clarifying}} \\
\hline
 GPT-4   & gpt-4            &  & \cellcolor{green!9}+0.08  & \cellcolor{green!16}-0.15 & \cellcolor{red!7}+0.07 &       &          0.59 &          2.26 &           1.32 &          0.00 &           0.00 \\
 GPT-4   & gpt-4            & \checkmark & \cellcolor{green!11}+0.11 & \cellcolor{green!16}-0.15 & \cellcolor{red!5}+0.04 &       &          0.61 &          2.45 &           1.26 &          0.00 &           0.00 \\
 GPT-3.5 & text-davinci-003 & & \cellcolor{green!6}+0.06  & \cellcolor{green!12}-0.12 & \cellcolor{red!6}+0.06 &       &          0.43 &          2.77 &           1.04 &          0.00 &           0.00 \\
 GPT-3.5 & text-davinci-003 & \checkmark & \cellcolor{green!14}+0.13 & \cellcolor{green!16}-0.16 & \cellcolor{red!3}+0.03 &       &          0.36 &          1.52 &           0.76 &          0.00 &           0.00 \\
\hline
\multicolumn{9}{l}{\emph{Filtering + Planning}} \\
\hline
 GPT-4   & gpt-4            &   & &      &      & \cellcolor{red!4}-0.03   &          2.97 &         50.01 &          18.19 &          0.00 &           0.00 \\
 GPT-3.5 & text-davinci-003 &   & &      &      & \cellcolor{red!16}-0.16  &          1.02 &         15.05 &           7.42 &          0.00 &           0.00 \\
\hline
\multicolumn{9}{l}{\emph{Filtering}} \\ \hline
 GPT-4   & gpt-4            &  &        &      &      & \cellcolor{green!2}+0.01 &          2.23 &         36.66 &          11.87 &          0.00 &           0.00 \\
 GPT-4   & gpt-4            &  \checkmark &        &      &      & \cellcolor{green!1}+0.00 &          2.14 &         38.39 &          12.71 &          0.01 &           0.00 \\
 GPT-3.5 & text-davinci-003 &  &        &      &      & \cellcolor{green!3}+0.02 &          1.59 &         19.74 &           7.61 &          0.01 &           0.00 \\
 GPT-3.5 & text-davinci-003 &  \checkmark &        &      &      & \cellcolor{green!1}+0.00 &          1.72 &         15.84 &           6.49 &          0.00 &           0.00 \\
\hline \hline
\multicolumn{9}{l}{\emph{Full Reasoning Chain (Clarifying, Filtering, Planning)}} \\ 
\hline
 GPT-4   & gpt-4            & \checkmark, - & \cellcolor{green!11}+0.11 & \cellcolor{green!16}-0.15 & \cellcolor{red!5}+0.04 & \cellcolor{green!2}+0.01 & 0.61 & 66.98 & 14.84 & 0.00 &           0.00 \\
 GPT-3.5 & text-davinci-003 & \checkmark, - & \cellcolor{green!14}+0.13 & \cellcolor{green!16}-0.16 & \cellcolor{red!3}+0.03 & \cellcolor{green!3}+0.02 & 0.40 & 33.61 & 11.73 & 0.01 &           0.00 \\
\hline
\end{tabular}
    
\caption{Summary of quantitative results for Sasha. Results are shown as the difference between the zero-shot approach in our empirical study relative to the measurement achieved by iterative components of Sasha's reasoning chain. In our ablation study, we test the effect of reasoning on full home templates (containing all devices and their settings) versus reasoning on simplified lists of devices (denoted by a \checkmark in the ``List'' column).}
\label{tab:sasha-quantitative-results}
\end{table}

We evaluate Sasha's performance against our home and command dataset from Section~\ref{feasibility-study} with 2 different LLMs in an ablation study. For repeatability, we set temperature $t=0.0$. Toward answering \textbf{RQ3}, our goal is to understand how different components of the reasoning chain influence the performance of the system, measured in terms of accuracy, false positives, false negatives, and relevance (defined in Section~\ref{sec:relevance-metrics}), all of which can be computed by simply examining the responses returned by Sasha.

\textbf{\em Result 1: Clarifying the goal separately from filtering devices reduces the rate of false positives.}
We compare the performance when combining Clarifying + Filtering into a single step versus Clarifying in a separate step. In the first case, we use a single prompt to ask the LLM to select any relevant devices, or to decline if no devices appear relevant (Clarifying + Filtering). In the latter case, we prompt the model to consider \emph{if} there are any relevant devices (i.e., Sasha's Clarifying component). In both cases, we measure the impact of reasoning on the full home template versus a simplified list of available devices. We report the results in Table~\ref{tab:sasha-quantitative-results}.

For both models, combined Clarifying + Filtering improves accuracy by reducing false positives. For GPT-3.5, reasoning on the list of devices provides a slight advantage; this is not the case for GPT-4. For GPT-4, the false negative rate slightly increases. The separate Clarifying step provides a larger performance benefit than combined Clarifying + Filtering in both cases. \aeone{This benefit is further increased when Clarifying is done on a list of devices.} Reduction in false positives is slightly offset by a small increase in false negatives---the Clarifying step, in essence, makes the output more conservative.

\textbf{\em Result 2: Filtering devices separately from planning the actions improves relevance.}
We evaluate the effect of combining Filtering + Planning (i.e., the model creates an action plan on the full set of devices) relative to separate Filtering and Planning steps (i.e., the model creates an action plan from the filtered set of devices) (Table~\ref{tab:sasha-quantitative-results}). For both models, combined Filtering + Planning reduces relevance over the zero-shot approach. When separating the Filtering step and reasoning on the full set of devices, we achieve a slight improvement in relevance for both models; when Filtering on a list of devices, performance is the same as the zero-shot method. 

\label{sec:sasha-results-summary}
\textit{\textbf{Summary.}} We draw the following conclusions from our quantitative evaluations of Sasha:

\begin{itemize}
    \item Clarifying reduces false positives for multiple models, with probable benefits to user satisfaction.~\textbf{\hyperref[rq3]{(RQ3)}}
    \item Filtering provides modest improvement to target device relevance for multiple models.~\textbf{\hyperref[rq3]{(RQ3)}}
\end{itemize}

\section{User Study}
\label{user-study}
\begin{figure}[t!]
    \centering
    \includegraphics[width=0.325\columnwidth]{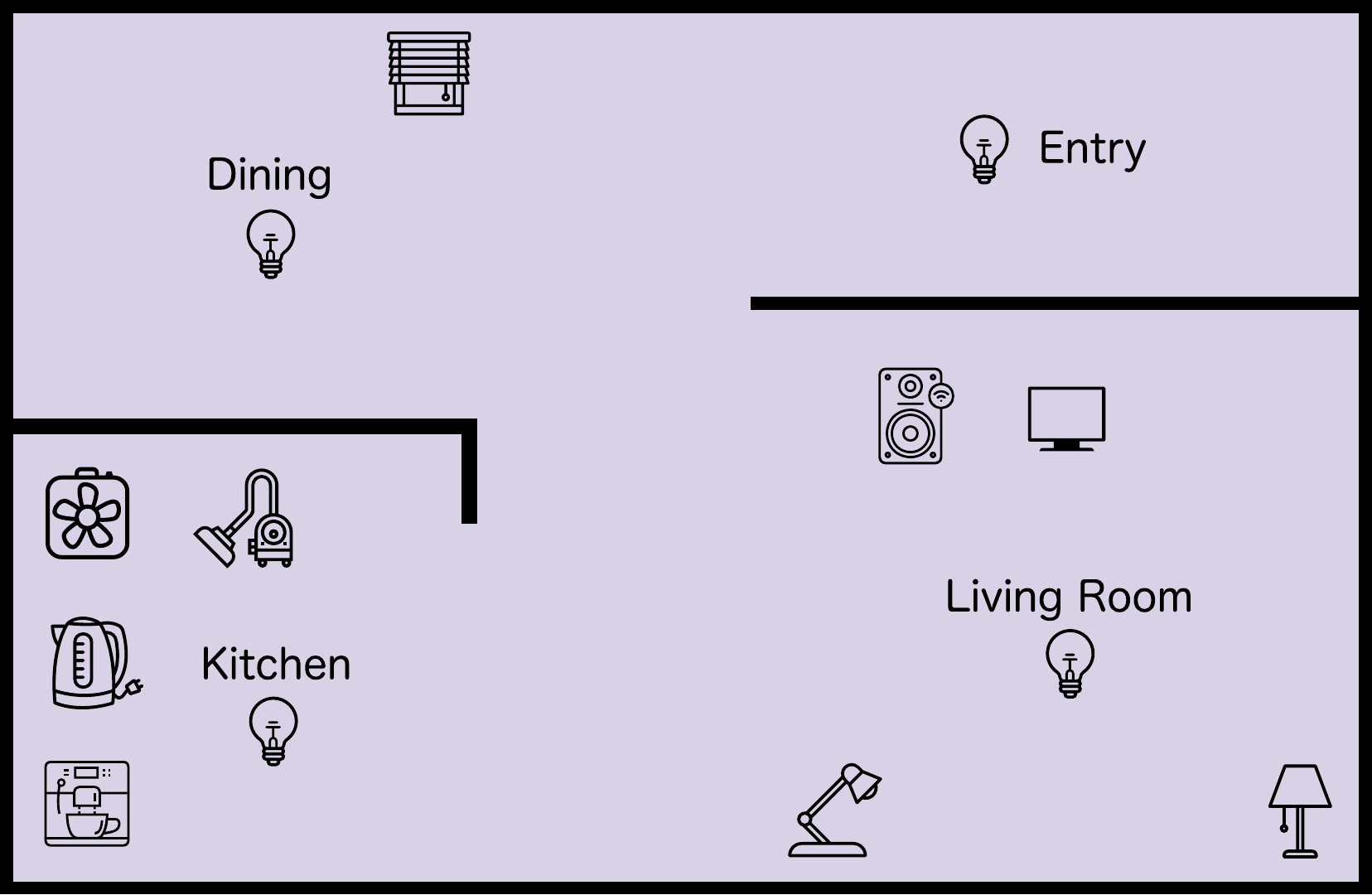} \hfill
    \includegraphics[width=0.325\columnwidth]{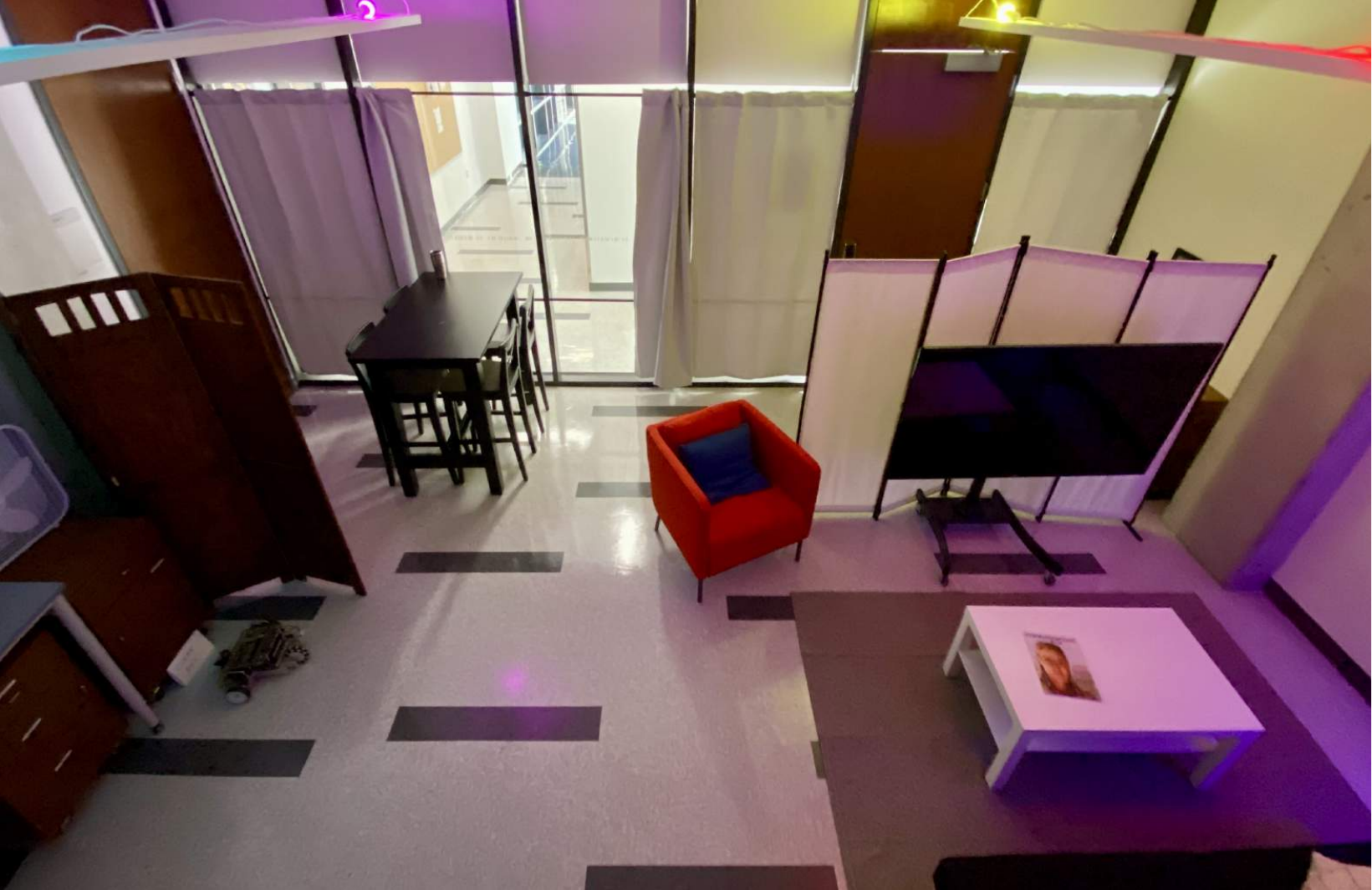} \hfill
    \includegraphics[width=0.325\columnwidth]{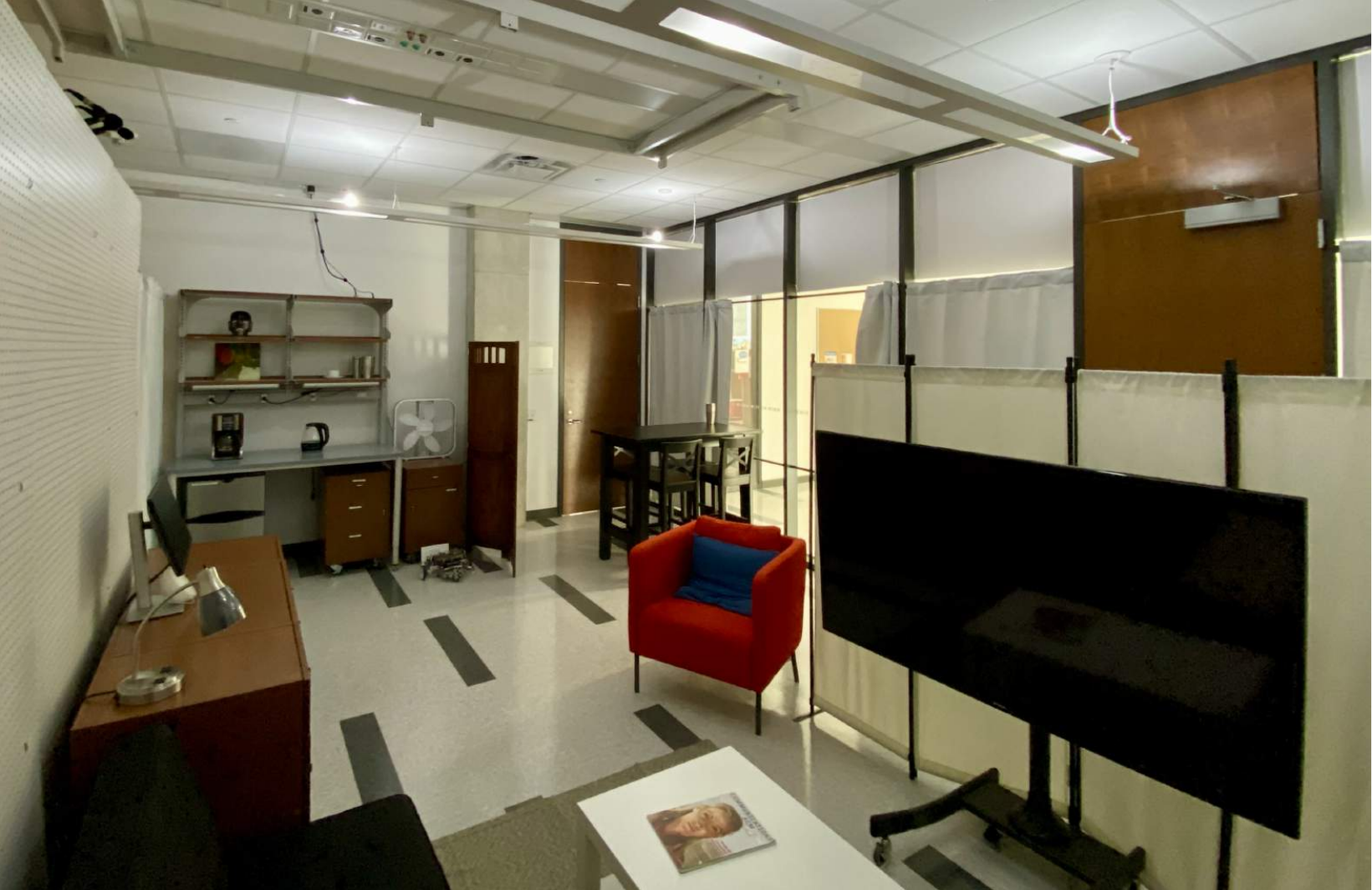}
    \caption{Test home for Sasha user study. We invite $N=7$ participants to give unconstrained commands to Sasha as they enact activities of daily living in the test home. The user study enabled us to discover the LLM-based smart home's capabilities and limitations when faced with user-generated scenarios.}
    \label{fig:sasha-floorplan}
\end{figure}

\aeone{Toward answering \textbf{RQ4}, we evaluate our implementation of Sasha in the presence of unpredictable user-generated scenarios} in a hands-on study. We constructed a test home in a large lab (Fig.~\ref{fig:sasha-floorplan}) containing smart lights, smart curtains, a coffee pot, an electric kettle, a fan, a smart speaker with 5 genres of music to choose from, a smart TV, and a ``cleaning robot''\footnote{Our cleaning robot lacks the ability to clean, but is illuminated by an indicator light when the system turns it on.}. We extended our Python implementation of Sasha to control these devices via several smart device APIs. We then invited $N=7$ participants to interact with Sasha in several open-ended scenarios of daily living. Since our sessions were time-constrained, we studied Sasha's responses to immediate goals only. 

\aetwo{\subsection{User Study Procedure \& Analysis}
\label{sec:sasha-user-study-procedure}}
\changed{\subsubsection{Participants and session structure} We conducted our user study with IRB oversight. \rfour{We recruited $N=7$ participants (Table~\ref{tab:lab-participants}) using email, snowball, and word-of-mouth.} Sessions were audio-recorded and lasted 30-45 minutes. During each session, a single participant (1)~received a 5 minute overview and demonstration of Sasha’s capabilities; (2)~enacted 3 preset scenarios of daily living while giving unconstrained, user-generated commands to the system; (3)~enacted custom scenarios related to the participant's personal routines; and (4)~participated in a closing interview. We define 3 preset scenarios (``Coming Home,'' ``Morning Routine,'' and ``Having Party''), each of which has several high-level goals (``Coming Home'': Have dinner, wind down, bedtime; ``Morning Routine'': Wake up, leave home; ``Having Party'': Prepare home, end party). 
After completing 3 preset scenarios, participants invented custom scenarios relating to their personal routines and gave commands accordingly. We instructed the participants to continue giving commands and feedback to Sasha until they felt that the devices in the room sufficiently supported each of the goals for that scenario. We instructed participants to use unconstrained language in commands and asked them to vocally share their perspectives on the system's actions. While Sasha is capable of interpreting spoken commands, for this controlled study, a researcher manually entered participants' commands. Since response latency is out of scope for our work, we instructed participants to share perspectives on the system's actions rather than system latency. The script for sessions is in Appendix~\ref{sec:appendix-session-script}.

\subsubsection{Qualitative analysis} \rfour{We gathered approximately 5 hours and 30 minutes of audio from our sessions, which we analyze to identify common themes of user interactions with Sasha.} We elicit instances where Sasha was able to support unconstrained user goals, as well as cases where the system's limitations were apparent. We label interactions based on the high-level scenario they occurred in, as well as the goal of the interaction.

\subsubsection{Quantitative analysis} During each session, we logged the state of Sasha's reasoning chain as users interacted with the system. We measure how frequently Sasha rejected commands, \rfour{the extent to which participants gave feedback,} and how many commands were used to meet each scenario's numerous goals.

\subsection{Observations}
\label{sec:sasha-user-study-observations}
\rfour{We present common themes from user interactions} in the following. Headers indicate the high-level scenario that the interaction occurred in, followed by the specific goal of the interaction.

\begin{table}
    \centering
    \begin{tabular}{llllll}
    \hline
       \textbf{ID} & \textbf{Gender}   & \textbf{Age}   & \textbf{Education}                      & \textbf{Smart Home Familiarity}   & \textbf{English Level}    \\
    \hline
 RWP1   & F        & 25-34 & Bachelor's                    & Moderately           & Native/bilingual \\
 RWP2   & F        & 55-64 & Bachelor’s 	                  & Slightly             & Native/bilingual \\
 RWP3   & F        & 18-24 & High school or equivalent	  & Very                 & Native/bilingual \\
 RWP4   & M        & 18-24 & Bachelor’s 	                  & Moderately           & Native/bilingual \\
 RWP5   & M        & 18-24 & Bachelor’s	                  & Slightly             & Native/bilingual \\
 RWP6   & M        & 18-24 & Graduate                      & Slightly             & Native/bilingual \\
 RWP7   & M        & 25-34 & Bachelor's                    & Very                 & Intermediate \\
    \hline
    \end{tabular}
    \caption{Demographic characteristics of real-world participants (RWP) in our user study of Sasha. RWP gave unconstrained commands during scenarios of daily living in our test home and provided perspectives on the system's performance.}
    \label{tab:lab-participants}
\end{table}

\emph{\textbf{Observation 1: A loosened command structure can support complex, multi-part goals.}} Participants took advantage of Sasha's loosened command structure to give commands that incorporated multiple steps:

\begin{tcolorbox}[colback=green!5,title={Morning Routine $\rightarrow$ Wake Up}]
\textbf{RWP4:} I'm too hot after my shower and I would like to get some cleaning done.

\textbf{Sasha:} I turned on the fan to help cool you down and started the cleaning robot to assist with cleaning.
\end{tcolorbox}

\begin{tcolorbox}[colback=green!5,title={User-Generated $\rightarrow$ Miscellaneous}]
\textbf{RWP7:} I want to have coffee at the dining room table and look outside.

\textbf{Sasha:} I have turned on the coffee machine in the kitchen to prepare your coffee. I have also opened the dining room curtains for you to look outside and adjusted the overhead light to a comfortable brightness.
\end{tcolorbox}

\begin{tcolorbox}[colback=green!5,title={User-Generated $\rightarrow$ Miscellaneous}]
\textbf{RWP5:} 
I'm done watching the match and I'm gonna have some coffee and my dinner.

\textbf{Sasha:} The TV in the living room has been turned off as you finished watching the match. The coffee machine in the kitchen has been turned on for your coffee. The overhead lights in the kitchen and dining room have been turned on with a brightness of 70 and a color of white for your dinner.

\textbf{RWP5:} Yeah, that makes sense.
\end{tcolorbox}






\noindent \emph{\textbf{Observation 2: Users leverage feedback to iteratively refine action plans.}} Participants often opened an interaction with an under-specified command, then refined action plans through iterative feedback. Their feedback was often equally under-specified, and Sasha was nonetheless able to incorporate it: 

\begin{tcolorbox}[colback=green!5,title={Having Party $\rightarrow$ End Party}]
\textbf{RWP4:} People are leaving, let's wind the party down.

\textbf{Sasha:} I have dimmed the lights in the living room, kitchen, and dining room to create a more relaxed atmosphere. The entry light is slightly brighter to guide people out. The music player in the living room is now playing ambient music to help wind down the party. The TV has been turned off.

\textbf{RWP4:} \textbf{(Feedback)} The music change was a bit jarring.

\textbf{Sasha:} Changed the music to a less jarring genre (folk) based on your feedback.

\textbf{RWP4:} Okay, this is much better. I like it... I wouldn't have thought of that.
\end{tcolorbox}

\noindent
In some cases, participants began an interaction with a completely under-specified command, then added more specificity through feedback:

\begin{tcolorbox}[colback=green!5,title={Coming Home $\rightarrow$ Bedtime}]
\textbf{RWP1:} I'm going to bed.

\textbf{Sasha:} All lights and TV in the living room and entry have been turned off for your bedtime.

\textbf{RWP1: (Feedback)} You also have to turn off the light in the kitchen.

\textbf{Sasha:} I have updated the action plan to also turn off the light in the kitchen as per your feedback.
\end{tcolorbox}

\begin{tcolorbox}[colback=green!5,title={Coming Home $\rightarrow$ Wind Down}]
\textbf{RWP3:} I'm going to relax now.

\textbf{Sasha:} I have turned off the reading lamp and TV in the living room to create a relaxing environment. I have also turned on the music player and set it to play ambient music.

\textbf{RWP3: (Feedback)} I wanted to turn on the TV.

\textbf{Sasha:} I have updated the action plan to turn on the TV as per your feedback.
\end{tcolorbox}

\noindent
In another case, Sasha's responses to feedback were less graceful. When a participant tried to use feedback to choose their preferred color of reading light, they got stuck in a back-and-forth with Sasha:

\begin{tcolorbox}[colback=red!5,title={Coming Home $\rightarrow$ Wind Down}]
\textbf{RWP2:} Can you change the reading lamp to a cooler tone?

\textbf{Sasha:} The reading lamp in the living room has been changed to a cooler tone (deep blue)

\textbf{RWP2:} [Laughs] \textbf{(Feedback)} That's too cool.

\textbf{Sasha:} Changed the color of the reading lamp to a warmer color (deep red) to address the user's feedback that the original color was too cool.

\textbf{RWP2:} Ah, wow. \textbf{(Feedback)} That's a little too red.

\textbf{Sasha:} I have reduced the red intensity and added some green and blue to the color to make it less red.

\textbf{RWP2:} Okay, uh. Hmm. A little pink.
\end{tcolorbox}

\noindent \emph{\textbf{Observation 3: Greater sensitivity to individual preferences and context is necessary.}} Some interactions highlighted a need for Sasha to learn individual preferences over time:

\begin{tcolorbox}[colback=red!5,title={Morning Routine $\rightarrow$ Leave Home}]
\textbf{RWP2:} I'm heading out Sasha, please make sure everything is turned off.

\textbf{Sasha:} All devices have been turned off as you are heading out. (cleaning robot remains on)

\textbf{RWP2:} Will something go on the fritz while I'm gone? I'd want that turned off.
\end{tcolorbox}

\noindent
Another participant expressed the opposite preference in the same scenario:

\begin{tcolorbox}[colback=green!5,title={Morning Routine $\rightarrow$ Leave Home}]
\textbf{RWP5:} I'm leaving for work.

\textbf{Sasha:} All lights, TV, and music player have been turned off. The cleaning robot in the kitchen has been turned on to clean while you're away.

\textbf{RWP5:} Yeah, in an ideal scenario, that's what I expect. To do the daily chores once I'm out.
\end{tcolorbox}

\noindent
One participant raised an interesting point about the risks posed by Sasha's lack of sensitivity to context:

\begin{tcolorbox}[colback=red!5,title={Having Party $\rightarrow$ End Party}]
\textbf{RWP1:} My friend threw up in the kitchen.

\textbf{Sasha:} I have turned on the overhead light in the kitchen to its maximum brightness for better visibility and activated the cleaning robot to clean up the mess.

\textbf{RWP1:} My immediate thought is, [that's] a good idea...Like, it made that connection...But, the limit on that is, like, that's not what the device can do...the device is not capable of cleaning vomit. So, in actual practice, that would just make a mess. I feel like it needs to know what the limits of the devices are.
\end{tcolorbox}

\noindent \emph{\textbf{Observation 4: Plans can introduce creative actions that the user may not think of themselves.}} In several cases, participants relied on Sasha to make decisions for them, or found that the system was able to craft plans that the participant would not have otherwise considered:

\begin{tcolorbox}[colback=green!5,title={Morning Routine $\rightarrow$ Wake Up}]
\textbf{RWP6:} I'm not sure if I want coffee or tea. Pick one for me.

\textbf{Sasha:} I have chosen coffee for you. The coffee machine is now on.
\end{tcolorbox}

\begin{tcolorbox}[colback=green!5,title={User-Generated $\rightarrow$ Improve Mood}]
\textbf{RWP6:} Help me make my girlfriend happy.

\textbf{Sasha:} I have set the music player in the living room to play an ambient playlist, started the coffee machine in the kitchen, and opened the curtains in the dining room to create a pleasant atmosphere.

\textbf{RWP6:} I think that would actually make her really happy. She likes coffee.
\end{tcolorbox}

\begin{tcolorbox}[colback=green!5,title={User-Generated $\rightarrow$ Having Visitors}]
\textbf{RWP4:} My parents are coming to visit.

\textbf{Sasha:} I have set the lights in the living room, entry, kitchen, and dining room to a warm brightness. The TV is off to encourage conversation. The music player in the living room is set to play a jazz playlist for a relaxing atmosphere. The coffee machine in the kitchen is on for your parents to enjoy a cup of coffee. The cleaning robot is activated to ensure the house is clean. The curtains are open to let in natural light.

\textbf{RWP4:} That one surprises me the most. The music choice and the coffee machine and the lights. I feel like it thought of everything in one command...But without me having to say all the steps.
\end{tcolorbox}

\noindent \emph{\textbf{Observation 5: Complex goals can be accomplished with only a few unconstrained interactions.}} We summarize quantitative results from our sessions in Appendix~\ref{sec:appendix-session-script}. Participants reported goals as accomplished in about 3 commands on average; this number varies, however, based on the complexity of the goal. Participants invoked feedback to refine action plans at a rate of 0.27 across 198 commands over all sessions. Feedback was used to incorporate user preferences and to correct oversights caused by the system's lack of knowledge of state (e.g., failing to target devices that had state set by prior interactions). We considered any rejected command to be a false negative. This was rare, at a rate of 0.06---typically when commands were completely under-specified, such as ``Goodbye, Sasha'' or ``Is it raining outside?''}

\textbf{\emph{Summary.}} \label{sec:user-study-summary} We make the following observations from our user study:

\begin{itemize}
    \item Feedback enables users to refine the aspects of action plans that are inherently subjective, but is only required in the minority of cases.~\textbf{\hyperref[rq3]{(RQ3)}}
    \item A loosened command structure enables the system to accomplish surprisingly complex, multi-part goals that were overlooked by our initial evaluation.~\textbf{\hyperref[rq4]{(RQ4)}}
    \item The system's effectiveness is limited in certain cases by insensitivity to user preferences and context.~\textbf{\hyperref[rq4]{(RQ4)}}
    \item The system's creative choices can help users accomplish their goals in few steps and with added flexibility, without a need for hardcoded routines.~\textbf{\hyperref[rq4]{(RQ4)}}
\end{itemize}

\section{Discussion}
\rthree{
\label{discussion}
Our results raise interesting questions for future research. We discuss these in the following.

\textbf{\em Defining a scope for future natural language interfaces in smart homes.}
A design challenge in natural language interfaces is to define their scope~\cite{mu2019we}, and make their limits apparent to users so that they can form a mental model of their capabilities~\cite{motta2022exploring}. While the system we have introduced significantly expands this scope, the same design challenge remains. Indeed, while current practical systems tend toward false negatives by rejecting commands that do not obey the command structure, an LLM-based system introduces a new risk of false positives by forming non-sensical plans in response to commands that are otherwise out of scope. In cases where security is a concern, e.g., in a home with smart locks, false positives introduce risks. We observed in our evaluations that strategies to reduce false positives tend to increase false negatives. A system that tends toward the former has greater creative capabilities with added risk, while a system with a tendency for the latter is close to the current state of the practice. The risk of false positives is lower when commands have evident goals but increases as users explore the capabilities of the system with different commands. This is common when users first acquire an assistant, or when children interact with smart assistants~\cite{garg2020he}. The question remains: \emph{how do we scope and design an LLM-based system to provide sufficient creative capabilities at minimal risk?}

\textbf{\em Long-term adaptation to user preferences and contexts.}
Sasha incorporates user feedback to improve action plans but does not provide long-term adaptation to user preferences or contexts. This is a desirable feature for smart assistants~\cite{reisinger2022user}, and our user study reiterates the need for it. In its simplest form, this adaptation entails a system learning that a user prefers, e.g., one genre of music over another when they give a command like ``help me relax''. At its most nuanced, however, an adaptive system would better understand the capabilities and contexts of individual devices---a cleaning robot, for instance, can clean, but different robots have different capabilities, and this context necessarily informs how the system reasons about them in relation to user goals. Further, the right action to meet a user's goal can vary depending on situational context: ``get the house ready for my friends'' at 12:00PM demands one set of actions, while the same command at 9:00PM demands a different set of actions. Our relevance score provides a coarse-grained notion of how well Sasha's design does at targeting relevant devices, but without the level of nuance that this particular problem will demand. User-provided feedback includes valuable information that could aid in this long-term adaptation. In general, determining \emph{how to adapt LLMs to individual smart home preferences and constraints} is an open challenge.

\textbf{\em Sensitivity to current state.}
We reason about user goals in relation to devices and assignable settings without sensitivity to their current state. Many goals, however, inherently depend on this state. A goal like ``make it less chilly in here'' expresses a desire to move from the current state to a new state; likewise a persistent goal like ``help me lower my power bill'' entails a more complex change in behavior from the current state (or set of states). We show in our evaluations that the system is able to generate surprisingly helpful and complex action plans, but suspect that in a longitudinal setting this sensitivity to state becomes necessary. If, for instance, a user asks the system to ``light it up a bit'' and the temperature parameter of the underlying model is set to $0.0$ (i.e., the output tokens of the action plan will always follow the most probable sequence), the system will always assign the same setting to the lights. This challenge of identifying \emph{relevant state} is beyond the scope of our work.
}

\section{Conclusion}
\label{conclusion}
We introduce the use of large language models (LLMs) to \emph{flexibly achieve user goals} using devices and sensors in smart homes. We framed the problem in terms of creating high-quality, executable ``action plans'' in JSON that either assign settings to available devices or propose triggers and actions for automation routines. Using a novel prototype system and empirical study, we performed qualitative and quantitative analysis of smart home action plans created by 4 different LLMs in response to 40 under-specified user commands in 3 homes. Based on our analysis of 600 responses from $N=20$ survey participants, we identified 7 failure modes that diminish user satisfaction, along with a set of corresponding metrics. The quality of action plans was hampered by LLMs' tendency to target irrelevant devices, assign improper settings, target extra devices, and choose suboptimal sensor triggers and values for automation routines. Toward addressing the issue of \emph{false positives} when targeting devices, along with a need to reduce subjectivity by incorporating user feedback, we proposed Sasha, a \underline{s}m\underline{a}rter \underline{s}mart \underline{h}ome \underline{a}ssistant. Sasha's novelty lies in an iterative reasoning process that decomposes the LLM's reasoning about the action plan into 4 components: clarifying the goal, filtering the devices, planning the action, and iteratively improving plans with user feedback. We justify each component of the process by ablation study, showing that Sasha's action plans have a significantly lower rate of false positives than the na\"{i}ve approach. Finally, we conduct a user study of Sasha with $N=7$ participants who interacted with the system during unconstrained scenarios, contributing a set of observations about how users goals may be supported by this new form of smart home.

\begin{acks}
We thank the anonymous reviewers for their constructive feedback on the paper. This work was funded in part by the National Science Foundation under grant CNS-1909221. Any opinions, findings, conclusions, or recommendations expressed are those of the authors and do not necessarily reflect the views of the NSF.
\end{acks}

\bibliographystyle{ACM-Reference-Format}
\bibliography{refs}
\newpage
\begin{appendices}
\section{User Commands}
\label{appendix-commands}
\begingroup
\renewcommand{\arraystretch}{.91}\small
\begin{tabularx}{\textwidth}{|X|l|r|}
    \hline
    \textbf{Command} & \textbf{Example IFTTT Routine}~\cite{yu2021analysis} & \textbf{Goal} \\ \hline \hline
    \multicolumn{3}{|l|}{{\it Ambient Temperature}} \\ \hline
    make it less chilly in here & ``Turn A/C off'' & immediate \\
    help me cool off  & ``Turn A/C on'' & immediate \\
    make it less stuffy in here	 & `Circulate air when CO$^2$ rises'' & immediate \\
    turn the AC off when it's cold outside	& ``Automatically turn off AC if it's cool outside'' & persistent \\
    turn the heat off when it's hot outside  & ``Warm weather...time to turn off the heating'' & persistent \\
    make it comfortable at night & ``Adjust the temperature... when I get into bed'' & persistent \\ \hline 
    \multicolumn{3}{|l|}{{\it Ambient Luminance}} \\ \hline
    make it less bright	& ``Change brightness of your light'' & immediate \\
    get ready for bed & ``Turn lights out for bed with Alexa and Hue'' & immediate \\
    help me see better & ``Turn on lights if it becomes dark'' & immediate \\
    keep it well-lit after sundown & ``Turn your lights on after sunset'' & persistent \\
    use natural light once the sun is up & ``Turn off lights at sunrise'' & persistent \\
    keep the lights off when I'm gone & ``Turn off lights when leaving a geofence'' & persistent \\ \hline 
    \multicolumn{3}{|l|}{{\it Security}} \\ \hline
    check the front door  & ``Close the garage door'' & immediate \\
    lock up & ``Tell Alexa to lock your door'' & immediate \\
    let the guest in & ``Ask Alexa to unlock door when guests arrive'' & immediate \\
    lock the door when I leave & ``When I leave my house, close garage door'' & persistent \\
    let me know when there's a visitor & ``Turn on lights when there is motion at door'' & persistent \\
    keep the home safe during the day & ``"Wake the camera up when I leave home'' &  persistent \\ \hline 
    \multicolumn{3}{|l|}{{\it Energy Saving}} \\ \hline 
    turn off the thermostat when I don't need it & ``If you exit an area, turn your heating off'' &  persistent \\
    save energy when I'm gone & ``Turn off TP-Link Plug when you leave area'' &  persistent \\
    help me save energy & ``Turn your lights off as you leave home'' &  persistent \\
    help me lower my power bill & ``Turn off TP-Link Plug when you leave area'' &  persistent \\ \hline 
    \multicolumn{3}{|l|}{{\it Ambient Atmosphere}} \\ \hline
    set up for a party & ``Start a party! Put your lights in disco mode'' &  immediate \\ 
    make it cozy in here & ``Create a warm atmosphere inside if it's cold'' &  immediate \\
    help me wind down & ``Change your Hue lights... to help you sleep'' &  immediate \\ 
    make it cozy when it rains & ``Warm the lights to a cozy... when rain starts'' &  persistent \\ 
    let me know when the weather is bad	& ``Alert me via Hue if the wind gets dangerous'' &  persistent \\
    help me sleep better & ``Make bedroom light orange at 10pm'' &  persistent \\ \hline 
    \multicolumn{3}{|l|}{{\it Robot Control}} \\ \hline
    clean up the bedroom & ``Tell Alexa to start your Roomba'' & immediate \\
    stop cleaning & ``Tell Google to turn off your bot'' &  immediate \\
    tidy up the house & ``Tell Alexa to start your Roomba'' &  immediate \\
    stop cleaning after sunset & ``Schedule cleaning at a certain time every day'' &  persistent \\
    don't clean when people are here & ``When I arrive home, dock Roomba'' &  persistent \\
    clean up while I'm gone	&  ``When I leave home, start a cleaning job'' & persistent \\ \hline 
    \multicolumn{3}{|l|}{{\it Other Appliances}} \\ \hline
    give me some privacy &  ``Turn on PowerView Schedules with Alexa'' & immediate \\
    let some sun in	& ``Turn on PowerView Schedules with Alexa'' & immediate \\
    finish the coffee & ``Say `Alexa trigger coffee' to start the coffee'' & immediate \\
    keep it humid at night & ``Set your air purifier to sleep mode at night'' & persistent \\
    I need coffee in the morning & ``Say `Alexa trigger coffee' to start the coffee'' & persistent \\
    let some sun in if the weather is nice & ``Turn on PowerView Schedules with Alexa'' & persistent \\ \hline
\end{tabularx}
\endgroup

\section{User Survey Design}
\label{appendix-survey}
\subsection{Demographic Questions}
\begin{figure}[H]
    \centering
    \includegraphics[page=1,width=0.85\columnwidth]{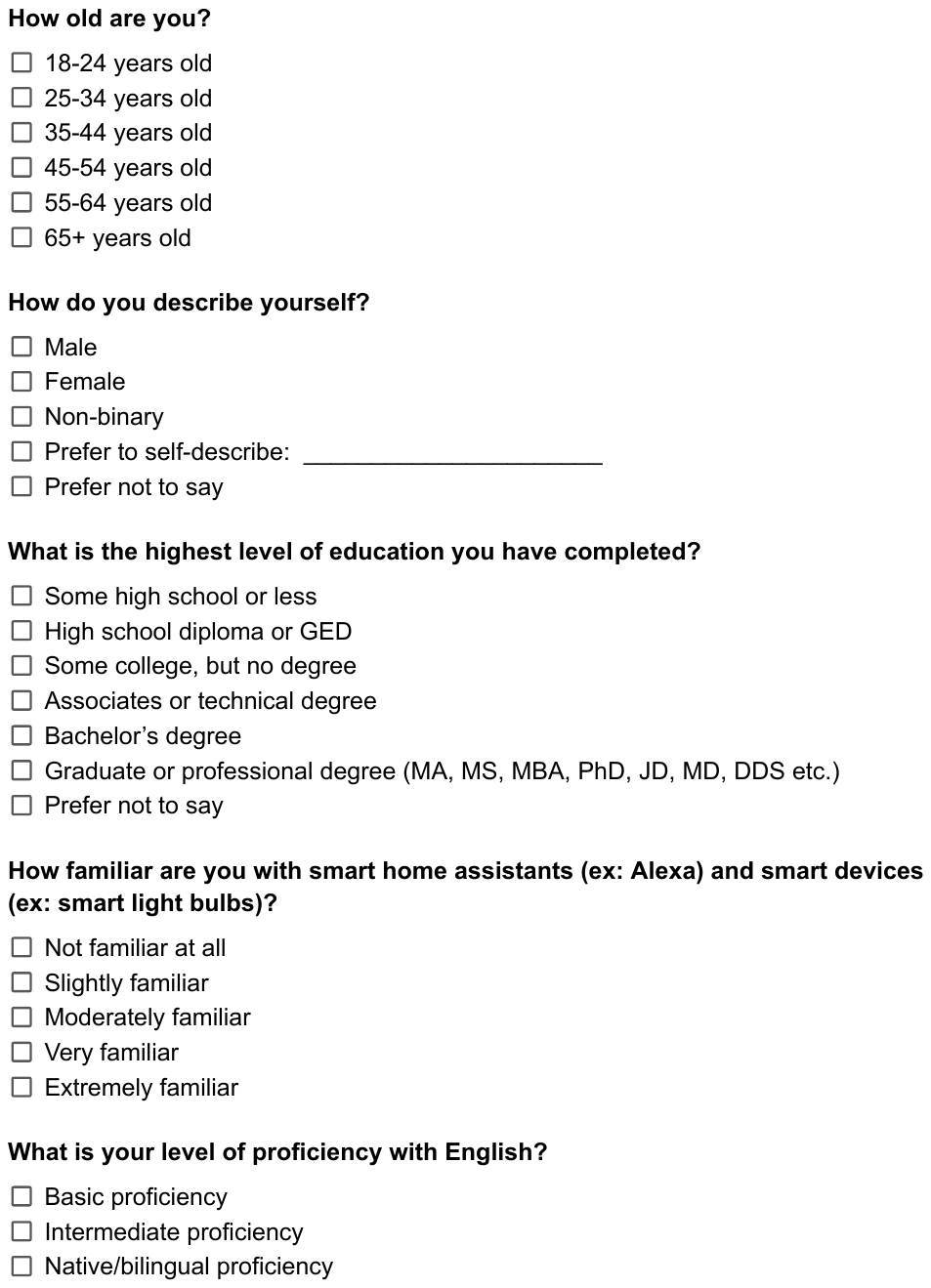}
\end{figure}

\subsection{Instructions}
\begin{figure}[H]
    \centering
    \includegraphics[page=1,width=0.88\columnwidth]{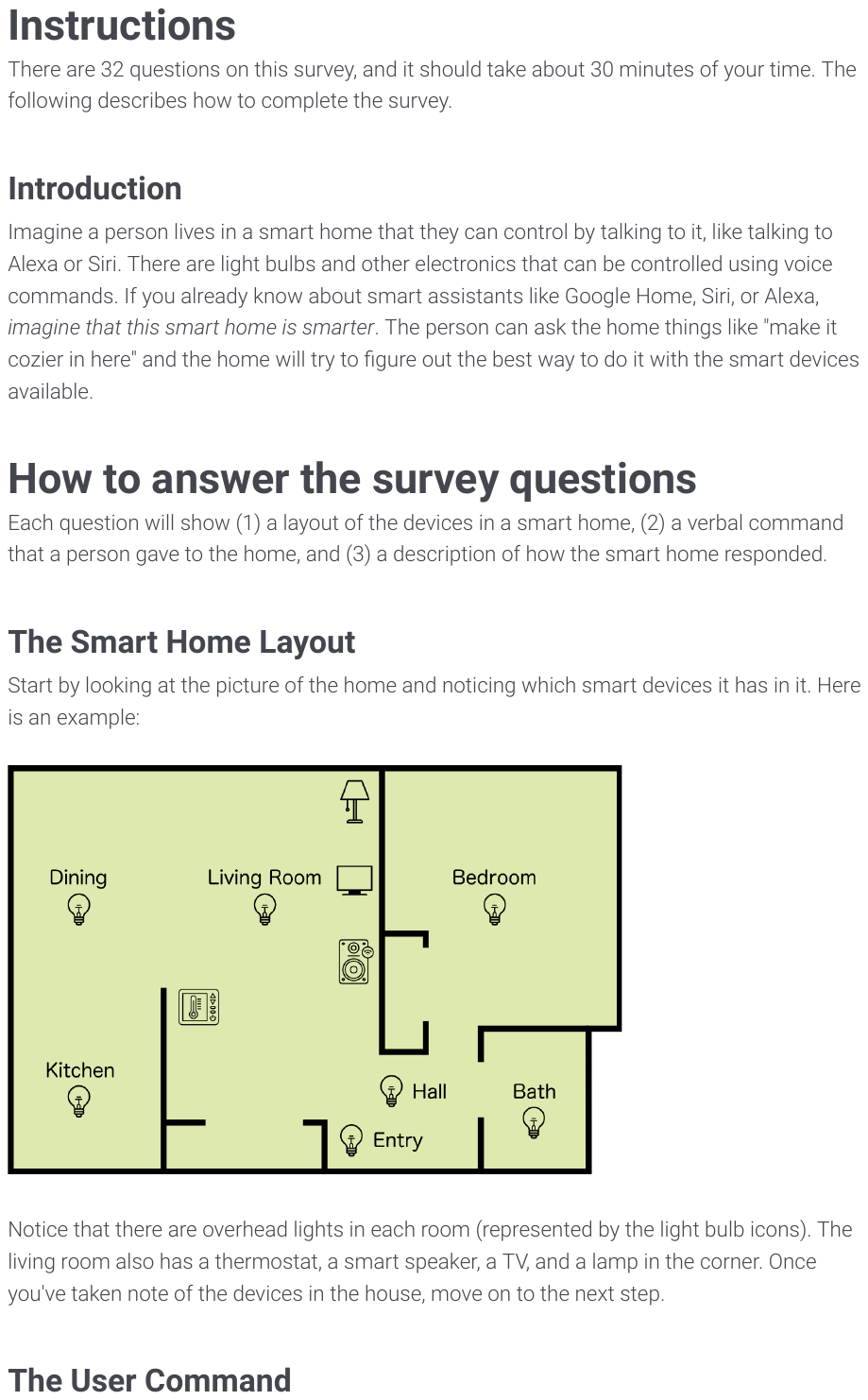}
\end{figure}

\begin{figure}[H]
    \centering
    \includegraphics[page=2,width=0.88\columnwidth]{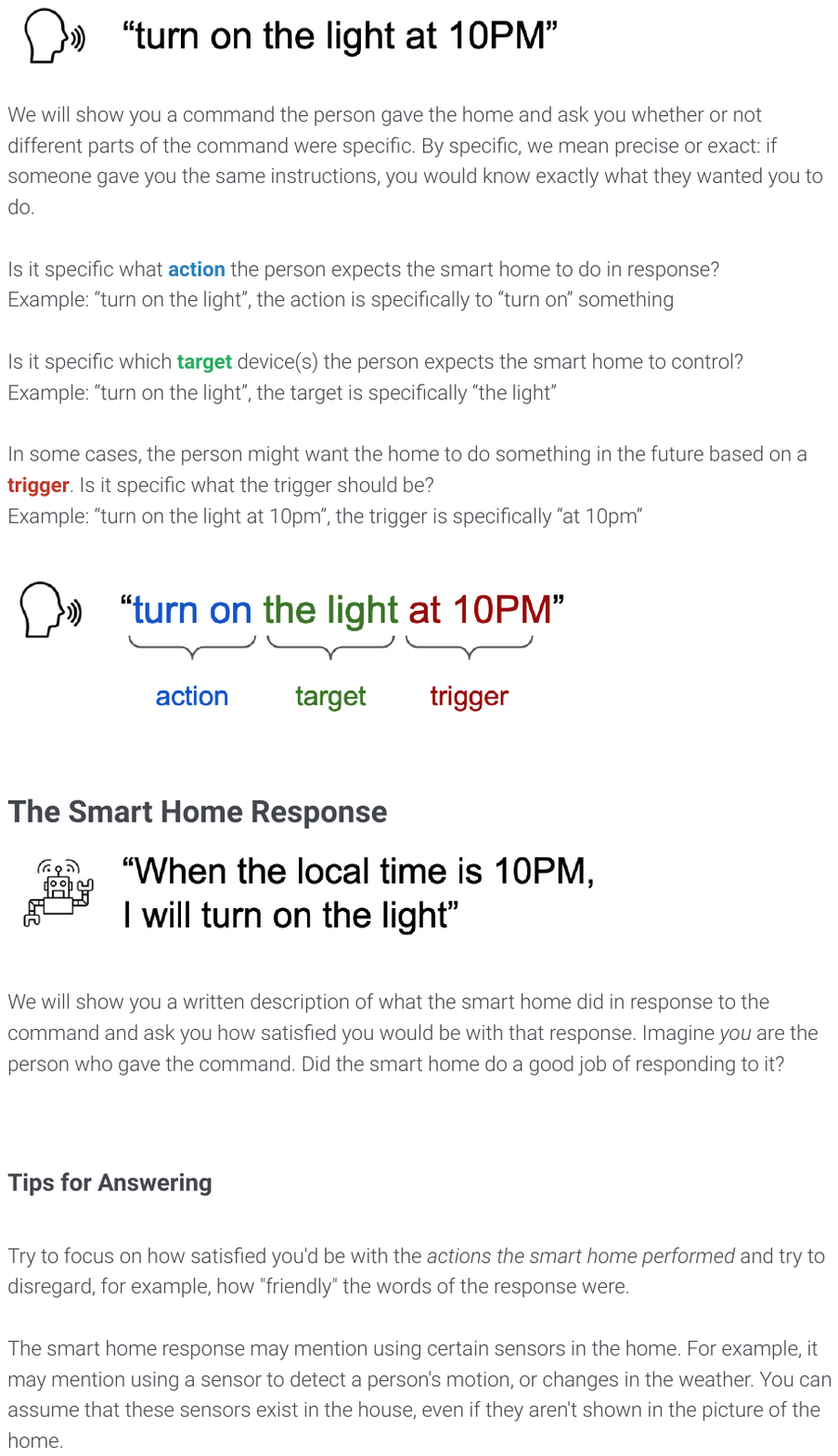}
\end{figure}

\begin{figure}[H]
    \centering
    \includegraphics[page=3,width=0.88\columnwidth]{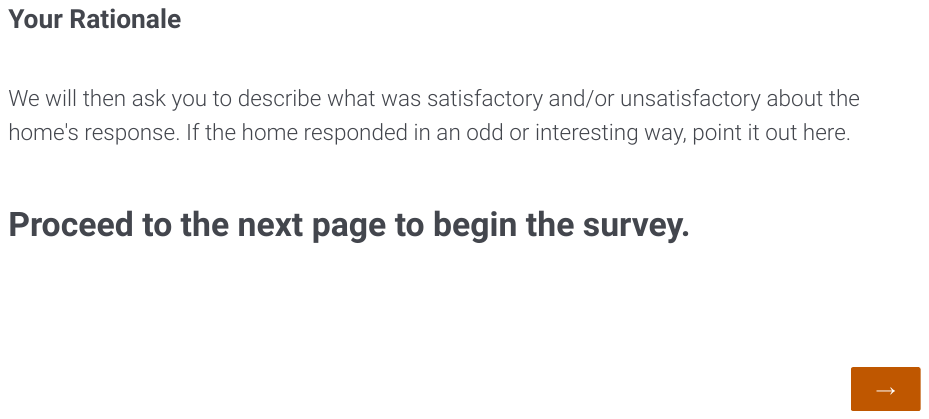}
\end{figure}

\subsection{Example Question}
\begin{figure}[H]
    \centering
    \includegraphics[page=1,width=0.88\columnwidth]{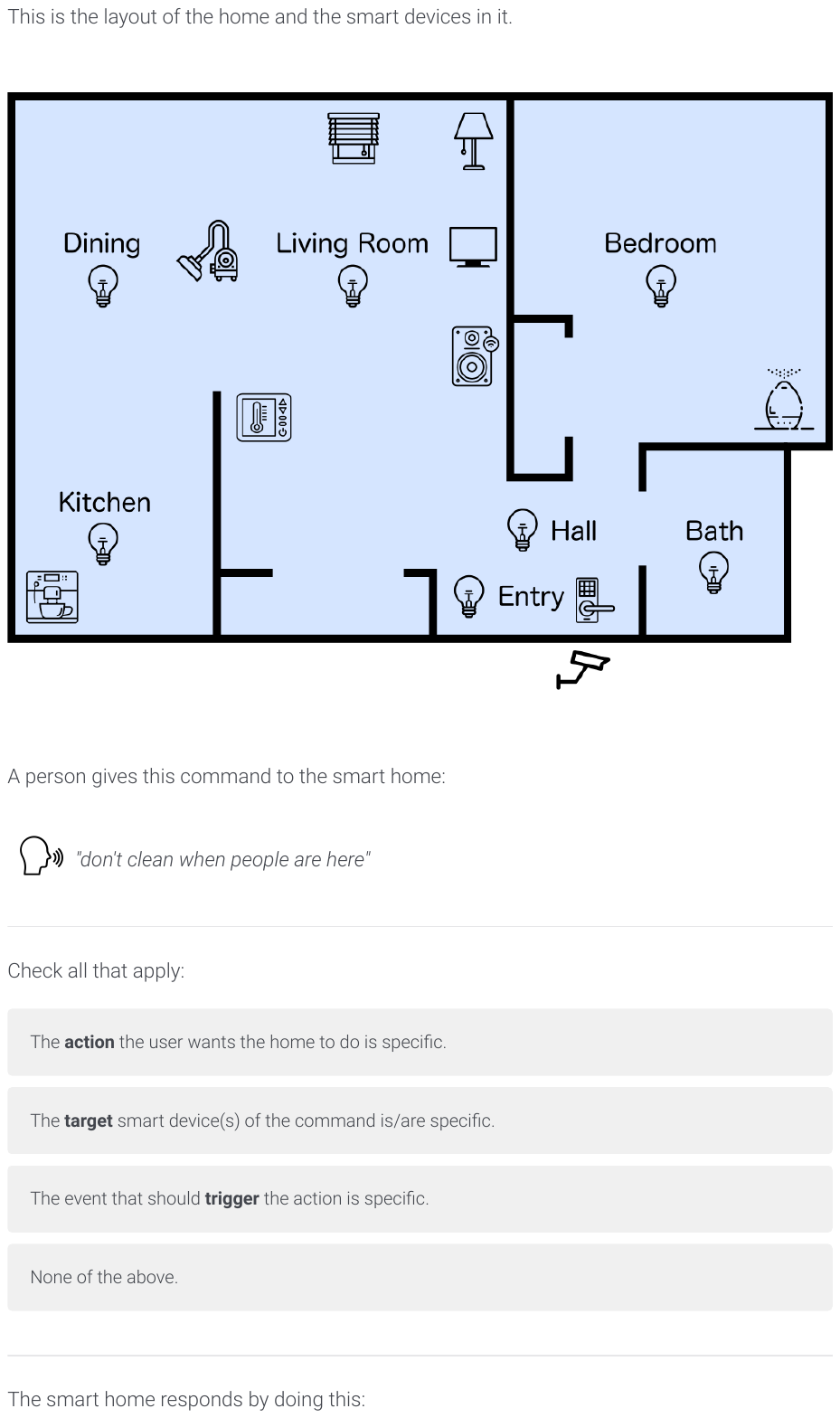}
\end{figure}

\begin{figure}[H]
    \centering
    \includegraphics[page=2,width=0.88\columnwidth]{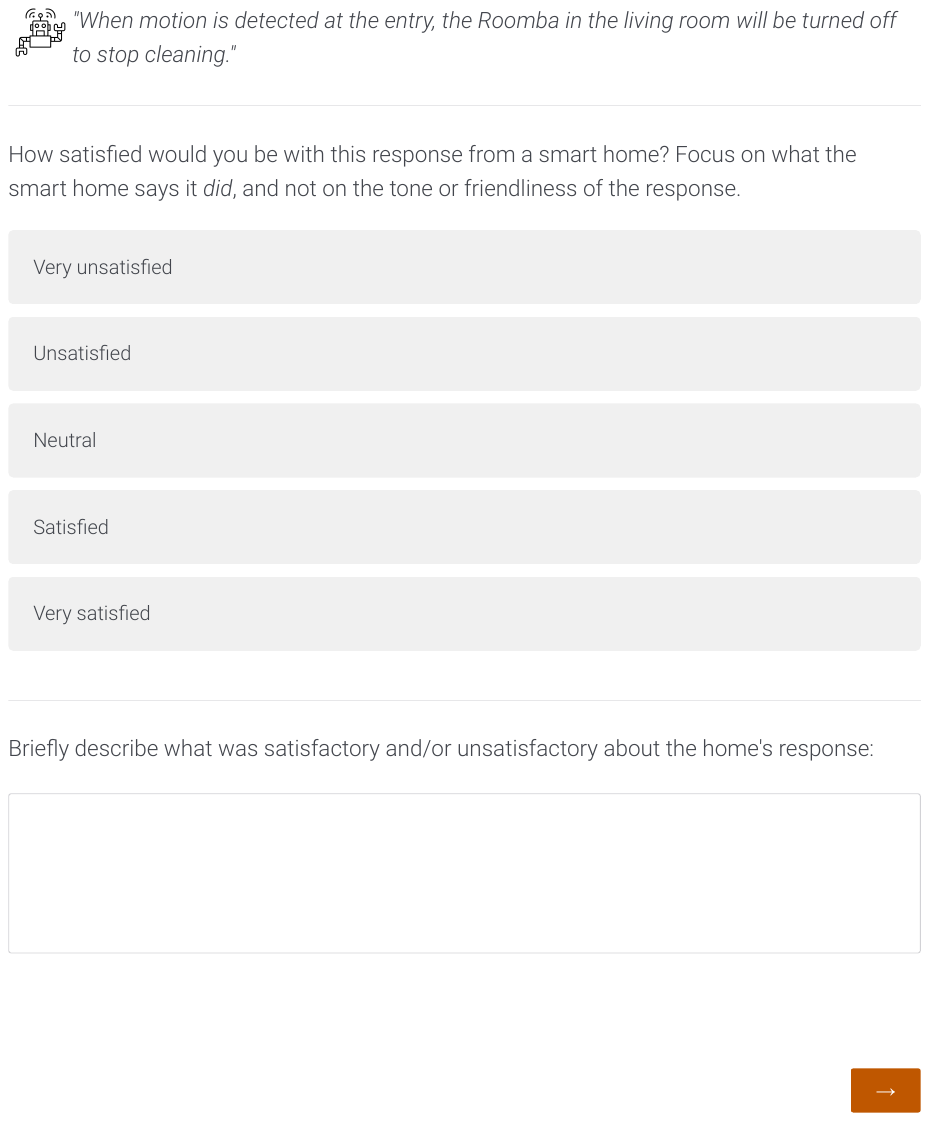}
\end{figure}

\begin{landscape}
\section{Failure Mode Examples}
\label{appendix-failure-modes}
\begin{tabular}{p{3.5cm}|p{4.5cm}lllp{4.5cm}}
\hline
 \textbf{Command}                                      & \textbf{Action Plan}                                                                                                                                                                                                                                                             & \textbf{Home}   & \textbf{Category}   & \textbf{Failure Mode}           & \textbf{Rationale}                                       \\
\hline
 stop cleaning after sunset                   & When the local time sensor detects that it is sunset, the smart speaker in the living room will be commanded to stop the cleaning process.                                                                                                                              & $h_2$     & Device          & No option exists       & SP18: Speakers do not clean                           \\ \hline
 make it less stuffy in here             & The thermostat in the living room has been set to 70 degrees to make the room less stuffy.                                                                                                                                                                                                          & $h_3$     & Device          & Option exists          & SP19: Turning on the diffuser would have helped too.                                                                                                                                                                                                                                                                                                                                                                                                   \\ \hline
 give me some privacy                         & All lights and audio devices have been turned off for privacy.                                                                                                                                                                   & $h_2$     & Device          & Extra                  & SP13: I would like the lights to stay on \\ \hline
 don't clean when people are here             & When motion is detected in the entry, the cleaning robot will be turned off.                                                                                                                                                                                            & $h_1$     & Device          & Hallucinated           & SP12: There is no robot                               \\ \hline
 make it cozy in here               & Adjusted the lights to a warm color and dimmed them to create a cozy atmosphere. The thermostat in the living room has been set to a comfortable temperature. The smart speaker volume has been lowered.                                                                                                                                                                   & $h_2$     & Device          & Setting                & SP7: I would instead of lowering speaker volume, set it to specific value, thus preventing it from becoming super quiet if initially speakers were already very low.                                                                                                                                                       \\ \hline

 help me sleep better               & When motion is detected in the bedroom, the system will turn off the lights in the bedroom and living room. The thermostat in the living room will be set to a comfortable 68 degrees. The smart speaker in the living room will be turned on at a low volume to provide white noise, which can promote better sleep. & $h_2$     & Sensor          & Suboptimal choice      & SP6: I think the problem is that it bases its actions on the motion detector.                               \\ \hline

 turn the heat off when it's hot outside      & When the weather is hot, heating system will be turned off.                                                                                                                                                                                                         & $h_1$     & Sensor          & Trigger value          & SP12: Hot would need to be defined                    \\
\hline
\end{tabular}
\end{landscape}

\section{User Study Materials}
\subsection{Session Script}
\label{sec:appendix-session-script}
\noindent \textbf{Introduction.}
Today we’re testing a system called Sasha. Sasha stands for “smarter smart home assistant”. Sasha tries to figure out how it can use different devices in the home to help you with your goals. It does this in response to open-ended or conversational verbal commands. If I tell Sasha to ``make it cozy in here'' or I tell Sasha ``hey, I want to read a book in the living room'', it tries to figure out what it can do with the devices in the room to help you, without you telling it specifically what you want it to do. Sasha has a few limitations:

\begin{itemize}
    \item Sasha can’t access information on the internet. 
    \item Sasha doesn't know where you are in the room. 
    \item Sasha doesn't have memory. Unless you give \emph{feedback} on something Sasha just did, it starts from scratch.
    \item Sasha is slow to respond. As we work with Sasha today, try to be forgiving of the delay and focus on the things Sasha does in response to your commands.
\end{itemize}

\noindent \textbf{Participant's Role.}
We are interested in your responses to what Sasha does. We are also interested in how you phrase your commands to Sasha. We'll act out a few different scenarios, you can give Sasha different commands to get the house ready, and give us your perspectives. If it did something weird, tell us, like ``I don’t really know why it turned on that light over there''. If it did something good or surprising, let us know. Any perspective you share, good, bad, or specific, is helpful.

\noindent \textbf{Tour.} Let me show you each of the devices that Sasha can control.
\begin{itemize}
    \item \emph{Entryway.} Sasha can control the overhead light.
    \item \emph{Dining room.} Sasha can control the overhead light, and also open the curtains.
    \item \emph{Kitchen.} Sasha can control the overhead light, and turn the coffee pot and tea kettle on and off. When Sasha turns on either of them, the light above them will turn on [demonstrate]. Sasha can control this fan. There is also a ``cleaning robot'' down here. It can't actually clean; but Sasha thinks it can. When Sasha turns on the robot, a light will turn on and illuminate the robot.
    \item \emph{Living room.} Sasha can turn the TV on and off, control the overhead light, the two lamps, and choose from a few different genres of music to play on the stereo.
\end{itemize}

\noindent \textbf{Examples.}
Here are some example commands [input these commands].

\begin{itemize}
    \item ``I'm going to read a book in the living room''
    \item ``I think I wanna watch a movie'' (Feedback) ``The lights are too bright for a movie.''
    \item ``Can you let some sun in?''
\end{itemize}

\noindent \textbf{Scenarios.}
We will work through three scenarios, each of which has a few goals. An example of a scenario is something like ``coming home from work'', and a goal would be ``have dinner''. At the start of each scenario, we will turn off all the devices in the room. Your task is to work with Sasha to get the home set up in a way that supports your personal definition of the goals in that scenario. When you feel like the home is ready, let me know and we will move on to the next scenario. Once we are done with the pre-defined scenarios, we will try some scenarios that are a part of your personal routine.

\noindent \textbf{Closing Questions.}
\begin{itemize}
    \item Was there anything you were surprised that Sasha was able to do? What was most surprising?

    \item Was there anything you felt like Sasha should have been able to do but couldn't? Did you notice anything that Sasha consistently got wrong?

    \item Are there any other devices you could imagine Sasha connected to? How would or could you use it if we had those devices?

    \item Would you use Sasha at home or at work? What would you use it for? Why wouldn't you use it?
\end{itemize}

\newpage

\begin{landscape}
\subsection{Session Summaries}
\bgroup
\setlength\tabcolsep{3pt}
\begin{table}[H]
\centering
\begin{tabular}{llrrr|lrrr}
\hline
 \textbf{ID}                    & \textbf{Scenario}        &   \textbf{{\underline G}oals} &   \textbf{{\underline C}mds} &   \textbf{{\underline C}/{\underline G}} & \textbf{Example Command \& Goal}                                                                    &   \textbf{Fdbk} &   \textbf{$FN$} &   \textbf{Err} \\
\hline
 \multirow{4}{*}{RWP1} & Coming Home     &       3 &      8 &        2.67 & I'm going to bed. (Bedtime)                                                                &     0.12 &   0.00 &   0.12 \\
                       & Morning Routine &       2 &      7 &        3.50 & Make sure the lights are on while while I'm gone. (Leave Home)                             &     0.14 &   0.43 &   0.00 \\
                       & Having Party    &       2 &      8 &        4.00 & My friends need to leave. (End Party)                                                      &     0.25 &   0.00 &   0.00 \\
                       & User-Generated  &       1 &      8 &        8.00 & I feel sad. Cheer me up. (Improve Mood)                                                    &     0.25 &   0.00 &   0.12 \\ \hline
 \multirow{4}{*}{RWP2} & Coming Home     &       3 &     12 &        4.00 & Clean up all the dishes. (Have Dinner)                                                     &     0.50 &   0.00 &   0.00 \\
                       & Morning Routine &       2 &      6 &        3.00 & Good morning Sasha, please wake up the house. (Wake Up)                                 &     0.17 &   0.00 &   0.00 \\
                       & Having Party    &       2 &      5 &        2.50 & Please put on the party music. (Prepare Home)                                              &     0.00 &   0.00 &   0.00 \\
                       & User-Generated  &       2 &      3 &        1.50 & I need to take a call and it's noisy. (Phone Call)                                         &     0.00 &   0.33 &   0.00 \\ \hline 
 \multirow{4}{*}{RWP3} & Coming Home     &       3 &      8 &        2.67 & I'm going to relax now. (Wind Down)                                                        &     0.25 &   0.00 &   0.00 \\ 
                       & Morning Routine &       1 &      2 &        2.00 & I'm ready to start the day. (Wake Up)                                                      &     0.50 &   0.00 &   0.00 \\
                       & Having Party    &       2 &      3 &        1.50 & Wind down the party. (End Party)                                                        &     0.33 &   0.00 &   0.00 \\
                       & User-Generated  &       1 &      2 &        2.00 & Let's have a movie night. (TV Time)                                                        &     0.50 &   0.00 &   0.00 \\ \hline
 \multirow{4}{*}{RWP4} & Coming Home     &       3 &      7 &        2.33 & I'm too scared about the movie I'm watching. (Wind Down)                                   &     0.29 &   0.00 &   0.00 \\ 
                       & Morning Routine &       2 &      5 &        2.50 & I'm leaving, see you later in the afternoon. (Leave Home)                                  &     0.20 &   0.00 &   0.00 \\
                       & Having Party    &       2 &      5 &        2.50 & People are leaving, let's wind the party down. (End Party)                                 &     0.60 &   0.00 &   0.00 \\
                       & User-Generated  &       5 &      8 &        1.60 & My favorite team is about to play. (TV Time)                                               &     0.12 &   0.25 &   0.00 \\ \hline
 \multirow{4}{*}{RWP5} & Coming Home     &       3 &      8 &        2.67 & Make me some coffee. (Have Dinner)                                                         &     0.12 &   0.00 &   0.00 \\ 
                       & Morning Routine &       2 &      7 &        3.50 & I'm leaving for work, make sure you clean the house. (Leave Home)                          &     0.43 &   0.00 &   0.00 \\
                       & Having Party    &       2 &      3 &        1.50 & Set the home for a party. (Prepare Home)                                                   &     0.00 &   0.00 &   0.00 \\
                       & User-Generated  &       1 &      6 &        6.00 & I'm done for the day, I'm going out. (Miscellaneous)                                       &     0.00 &   0.17 &   0.00 \\ \hline
 \multirow{4}{*}{RWP6} & Coming Home     &       3 &     11 &        3.67 & I'm going to need a relaxing drink. (Have Dinner)                                          &     0.36 &   0.00 &   0.00 \\ 
                       & Morning Routine &       2 &      6 &        3.00 & Goodbye Sasha. (Leave Home)                                                                &     0.17 &   0.17 &   0.00 \\
                       & Having Party    &       2 &     12 &        6.00 & Set the mood for a party. (Prepare Home)                                                   &     0.25 &   0.08 &   0.00 \\
                       & User-Generated  &       2 &     11 &        5.50 & Help me make my girlfriend happy. (Improve Mood)                                          &     0.73 &   0.00 &   0.00 \\ \hline
 \multirow{4}{*}{RWP7} & Coming Home     &       3 &     10 &        3.33 & I need some hot water as well. (Wind Down)                                                 &     0.20 &   0.00 &   0.00 \\ 
                       & Morning Routine &       2 &     11 &        5.50 & I'm still feeling tired. (Wake Up)                                                         &     0.18 &   0.09 &   0.09 \\
                       & Having Party    &       2 &      6 &        3.00 & Set the lights in party mode. (Prepare Home)                                               &     0.17 &   0.00 &   0.00 \\
                       & User-Generated  &       2 &     10 &        5.00 & I want to have soup. (Mealtime)                                                            &     0.30 &   0.10 &   0.00 \\
\hline
\multicolumn{3}{r}{} & \textbf{198} & \textbf{3.39} & & \textbf{0.27} & \textbf{0.06} & \textbf{0.02} \\
\hline
\end{tabular}
    \caption{Quantitative summary of user study sessions. Each real-world participant (RWP) gave unconstrained commands during 3 preset scenarios with several goals, followed by user-generated scenarios. All commands were unconstrained; we provide examples. Goals: number of goals in the scenario, Cmds: total number of commands for all goals in the scenario, Fdbk: portion of commands providing Feedback, $FN$: false negatives, Err: JSON error rate.}
    \label{tab:sasha-user-study-quantitative-summary}
\end{table}
\egroup
\label{sec:appendix-session-summaries}
\end{landscape}

\end{appendices}
\end{document}